\documentclass[fleqn,usenatbib]{mnras}
\usepackage{newtxtext,newtxmath}
\usepackage[T1]{fontenc}
\usepackage{ae,aecompl}

\usepackage{graphicx} 
\usepackage{amsmath,amssymb,times}
\usepackage{deluxetable}
\usepackage{longtable}
\usepackage{epstopdf}
\usepackage[usenames,dvipsnames]{color}
\usepackage{float}
\usepackage[export]{adjustbox}
\usepackage{ulem}
\usepackage[export]{adjustbox}
\usepackage{hyperref}
\usepackage{pdflscape}
\hypersetup{
  colorlinks= true, 
  urlcolor  = Blue, 
  filecolor=black,
  runcolor=blue,
  linkcolor = red, 
  citecolor = blue 
} 
\newcommand{\kms}{km\,s$^{-1}$}
\newcommand{\nai}{Na\,{\sc I}}
\newcommand{\RN}[1]{%
  \textup{\uppercase\expandafter{\romannumeral#1}}%
}

\makeatletter
\newcommand{\NII}{N\,{\scriptsize II}}
\newcommand{\OIII}{O\,{\scriptsize III}}
\newcommand{\OI}{O\,{\scriptsize I}}

\newcommand{\HI}{H\,{\scriptsize I}}

\newcommand{\nad}{Na\,{\scriptsize D}}
\newcommand{\Hii}{H\,{\scriptsize II}}
\makeatother

\makeatletter
\def\@to{to}
\makeatother

\title[Resolved Outflows in Star-forming Galaxies]{Outflows in Star-forming Galaxies: Stacking Analyses of Resolved Winds and the Relation to Their Hosts' Properties}

\author[Roberts-Borsani et al.]{G.~W. Roberts-Borsani$^{1,2}$\thanks{E-mail: guidorb@astro.ucla.edu}, A. Saintonge$^{2}$, K.~L. Masters$^{3}$ and D.~V. Stark$^{3}$
\\
$^{1}$Department of Physics and Astronomy, University of California, Los Angeles, 430 Portola Plaza, Los Angeles, CA 90095, USA
\\
$^{2}$Department of Physics and Astronomy, University College London, Gower Street, London WC1E 6BT, UK \\
$^{3}$Department of Physics and Astronomy, Haverford College, 370 Lancaster Ave, Haverford, PA 19041, USA}

\date{Accepted XXX. Received YYY; in original form ZZZ}
\pubyear{2019}

\begin{document}
\label{firstpage}
\pagerange{\pageref{firstpage}--\pageref{lastpage}}
\maketitle

\begin{abstract}
Outflows form an integral component in regulating the gas cycling in and out of galaxies, although their impact on the galaxy hosts is still poorly understood. Here we present an analysis of 405 high mass (log M$_{*}$/M$_{\odot}\geqslant10$), star-forming galaxies (excluding AGN) with low inclinations at $z\sim$0, using stacking techniques of the \nad\ $\lambda\lambda$5889,5895 \AA\ neutral gas tracer in IFU observations from the MaNGA DR15 survey. We detect outflows in the central regions of 78/405 galaxies and determine their extent and power through the construction of stacked annuli. We find outflows are most powerful in central regions and extend out to $\sim$1R$_{e}$, with declining mass outflow rates and loading factors as a function of radius. The stacking of spaxels over key galaxy quantities reveals outflow detections in regions of high $\Sigma_{\text{SFR}}$ ($\gtrsim$0.01 M$_{\odot}$yr$^{-1}$kpc$^{-2}$) and $\Sigma_{M_{*}}$ ($\gtrsim$10$^{7}$ M$_{\odot}$kpc$^{-2}$) along the resolved main sequence. Clear correlations with $\Sigma_{\text{SFR}}$ suggest it is the main regulator of outflows, with a critical threshold of $\sim$0.01 M$_{\odot}$yr$^{-1}$kpc$^{-2}$ needed to escape the weight of the disk and launch them. Furthermore, measurements of the H$\delta$ and D$_{n}$4000 indices reveal virtually identical star formation histories between galaxies with outflows and those without. Finally, through stacking of \HI\ 21 cm observations for a subset of our sample, we find outflow galaxies show reduced \HI\ gas fractions at central velocities compared to their non-detection control counterparts, suggestive of some removal of \HI\ gas, likely in the central regions of the galaxies, but not enough to completely quench the host.
\end{abstract}

\begin{keywords}
galaxies: evolution -- galaxies: starburst -- ISM: jets and outflows -- ISM: inflows
\end{keywords}

\section{Introduction}


The evolution of galaxies is dictated primarily by the availability and regulation of cold gas available for star formation, both at low and high redshift \citep{saintonge13}. In this scenario, a galaxy fills its gas reservoir through accretion, converts the gas into stars over a free fall time, and ejects the subsequently metal-enriched gas out of the galaxy disk \citep{dave12,lilly13}. Such a picture is consistent with recent molecular and atomic gas observations of representative galaxies at $z\sim$0, which show significantly elevated gas fractions and star formation efficiencies for star-forming galaxies along and above the so-called ``star-forming main sequence'' compared to their passive counterparts \citep{saintonge17,catinella18}.

However, the conversion of gas into stars is an inefficient process, with scaling relations showing that only $\sim$1-10\% of the gas gets converted into stars at a given time \citep{kennicutt98,usero15,bigiel16}. Galactic-scale outflows, launched through the combined energy and momentum delivered by intense star formation or a supermassive black hole (an active galactic nucleus; AGN) are thought to contribute to this inefficiency through their potential to regulate and/or halt the star formation in the disks of galaxies, thereby providing the feedback necessary to deliver galaxies from the star-forming main sequence to the red cloud of passive galaxies. AGN feedback is thought to be the dominant regulating process for high mass galaxies at later times, whilst starburst-driven outflows are thought to dominate low mass galaxies at early times.

However, outflows are still poorly understood. Recent efforts aimed at understanding them have made progress by placing constraints on their prevalence, integrated properties, and the underlying correlations with their galaxy hosts through observations of relatively large samples of galaxies \citep{rupke05b,chen10,sugahara17,rb18}. Such studies have found that $\sim$100-1000 \kms\ outflows are common at all epochs among star-forming systems and AGN \citep{veilleux05,rupke05a,feruglio10,chen10,coil11,cicone16,cazzoli16,rb18} and particularly at high mass (log M$_{*}$/M$_{\odot}>10$) for normal galaxies of the local Universe along the main sequence when traced in absorption \citep{chen10,rb18,concas17,sugahara17}. Estimates of a mass outflow rate in star-forming galaxies, whilst extremely challenging and uncertain, have revealed values similar to the SFR of the galaxy, suggesting that more intense star formation expels higher quantities of gas \citep{rupke05b,rubin14,cazzoli16,rb18}. Similar correlations are found between the mass outflow rates and AGN luminosities for galaxies with an active nucleus, suggesting more extreme AGN expel gas at significantly higher rates \citep{sturm11,veilleux13,cicone14,ga17,fluetsch18}.

Such constraints have proved extremely valuable, however they are generally based on spectroscopic observations with limited and incomplete spatial coverage (e.g., long-slit or 3$''$ fiber) and as such may not be representative of entire galaxy systems. The need for more detailed and representative galaxy observations has seen the advent of large integral field unit (IFU) spectroscopic surveys such as CALIFA \citep{sanchez12}, SAMI \citep{croom12} and MaNGA \citep{bundy15}, which have heralded in a new era for statistical studies of representative galaxies and their outflows at parsec (pc) and kiloparsec (kpc) resolution. The arrival of such large IFU surveys, in conjunction with powerful instruments such as VLT/MUSE, have added valuable constraints on the prevalence of ionised-gas outflows in normal, individual galaxies, their driving properties, star-formation histories and AGN contributions on a per-spaxel basis (e.g., \citealt{ho-i16,lopezcoba19,rodriguezdelpino19}). Despite such progress, several outstanding questions remain: whether outflows can quench their hosts, either entirely or in specific regions, remains an open question and the primary driving properties in star-forming galaxies has yet to be fully determined.

In this paper, we expand on recent IFU work by using one of the largest IFU samples to date of star-forming galaxies observed as part of the MaNGA survey containing signatures of neutral gas (\nad) outflows, with the goal of determining through stacking techniques (i) the radial extent of outflows and their power, (ii) the main driving mechanism of outflows and the \nad\ absorption doublet, and (iii) the average star-formation histories and \HI\ gas fractions of outflow galaxies compared to those without. The paper is organised as follows. In \S2 and \S3 we present our sample selection, relevant data sets and stacking techniques using the MaNGA DR15 survey, respectively. \S4 presents an analysis of the radial extent of outflows and their power, whilst \S5 and \S6 explore the primary driving properties of outflows. We present the average \HI\ gas fractions of outflow galaxies and a control sample of galaxies without outflows in \S7. Finally, we provide a discussion and the star formation histories of outflow galaxies in \S8 and present our conclusions in \S9. Throughout the paper we adopt a $\Lambda$CDM cosmology with \textit{H}$_{0}$=70 km/s/Mpc, $\Omega_{m}=$0.3, and $\Omega_{\wedge}=$0.7, and assume a Chabrier IMF.

\section{Data and Sample}
\label{sec:datasets}
\subsection{The SDSS-IV/MaNGA Survey}
The SDSS-IV/Mapping Nearby Galaxies at APO (MaNGA; \citealt{bundy15}) survey is an IFU survey of the local Universe with the aim of obtaining spatially-resolved spectra for an unbiased sample of $\sim$10,000 galaxies from the NASA-Sloan Atlas (NSA) catalog, with log\,M$_{*}$/M$_{\odot}\gtrsim$9 at 0.01\,$\leqslant$\textit{z}$\leqslant$0.15 by 2020. Situated on the SDSS 2.5m telescope at Apache Point Observatory, the MaNGA instrument offers 29 differently-sized IFUs, each of which consists of a set of optical fibers grouped to form a hexagon and fed into the two BOSS spectrographs \citep{smee13}. These fiber bundles are split into 17 sets which can be used to observe chosen targets at a given time, with the remaining 12 used for flux calibration and additional single fibers for sky subtraction \citep{drory15}. The bundles cover up to 1.5\,$R_{e}$ and 2.5\,$R_{e}$ for the targeted galaxy sample and the resulting spectra have a wavelength coverage of 3600-10,400\,\AA\ with spectral resolution \textit{R}$\sim$2000 \citep{wake17}. 
On 10th December 2018, the MaNGA survey released its 15th Data Release (DR15; \citealt{aguado19}), along with data products such as maps of derived galaxy properties and continuum fits from the Data Reduction Pipeline (DRP; \citealt{law16}) and Data Analysis Pipeline (DAP; \citealt{westfall19}), as well as the Marvin Python package \citep{cherinka19} to view and analyse them. Additionally, the release also comes with new value added catalogs of spatial and integrated properties such as the Pipe3D \citep{sanchez18} and \HI\ follow up observations \citep{masters19}. In this study we make use of this release, which comprises completed observations for 4,656 galaxies.

\subsection{Sample Selection}
\label{subsec:mangasamp}
\begin{figure}
\center
 \includegraphics[width=0.9\columnwidth]{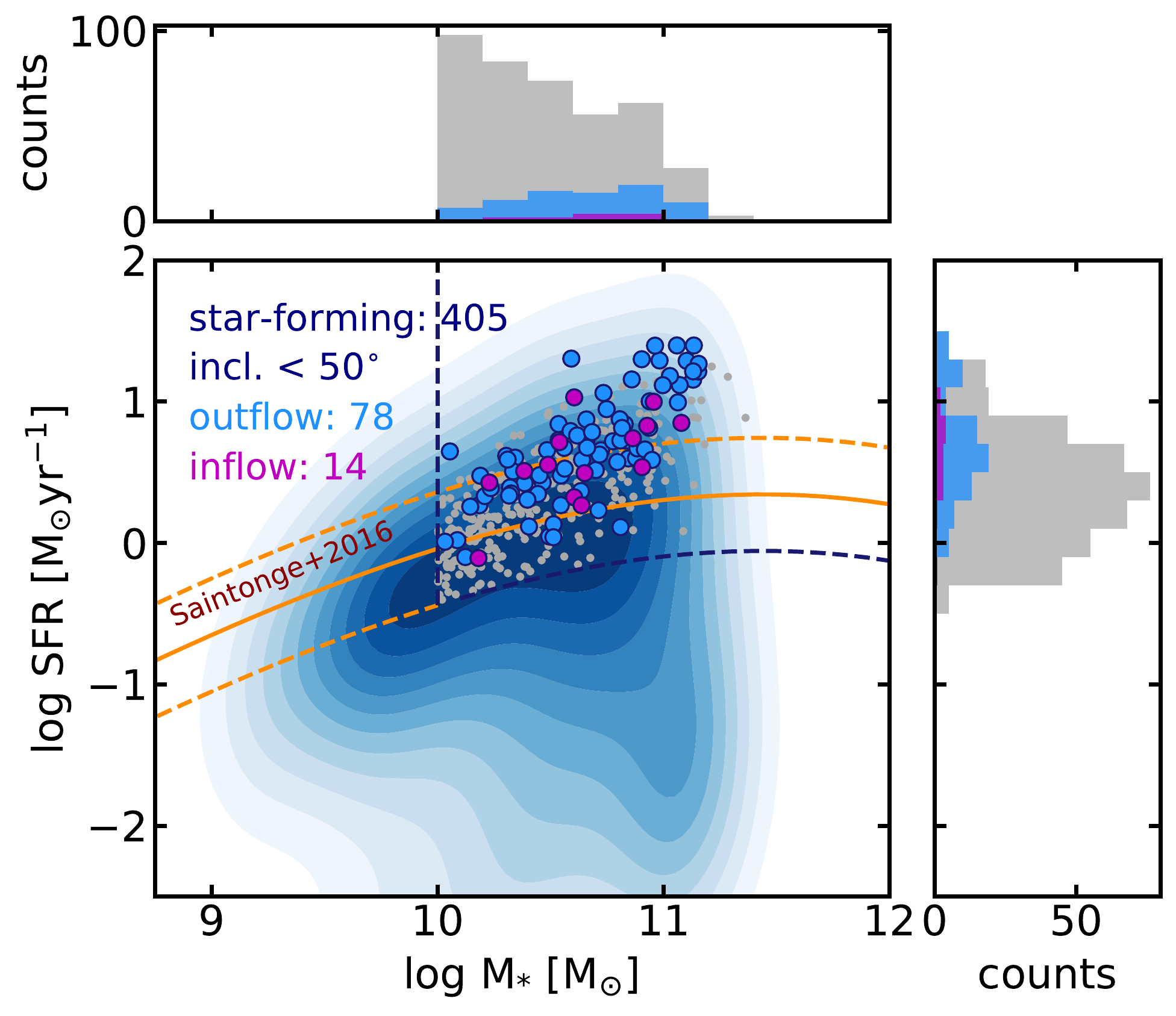}
  \caption{The SFR-M$_{*}$ plane and density contours of the full MaNGA DR15 sample and selected sample for this study. Gray dots represent the 405 galaxies found using our selection criteria (blue dashed lines), whilst the blue points mark galaxies found to have outflows in their central regions and magenta points galaxies found to have inflows. The orange solid and dashed lines mark the star-forming main sequence defined by \citet{saintonge16} and its lower and upper limit. Histograms of the selected galaxies' SFRs and stellar masses are shown to the top and right sides of the main plot.}
 \label{fig:sample}
\end{figure}

Motivated by recent results from \citet{rb18}, we begin our selection by performing cuts on galaxy stellar mass (log M$_{*}$/M$_{\odot}>$10), SFR (log SFR/M$_{\odot}$\,yr$^{-1}>$-2.332\,(log SFR/M$_{\odot}$\,yr$^{-1}$) + 0.4156\,(log SFR/M$_{\odot}$\,yr$^{-1}$)$^{2}$ - 0.01828\,(log SFR/M$_{\odot}$\,yr$^{-1}$)$^{3}$ - 0.4, corresponding to a rough lower limit of the galaxy main-sequence defined by \citealt{saintonge16}) and inclination ($i\leqslant$50$^{\circ}$, derived from the galaxy's $r$-band axis ratio, \textit{b}/\textit{a}) - three key galaxy properties known to influence the detection rates of neutral gas outflows. The stellar masses and SFRs used for this selection are taken from the Pipe3D catalog, which for each galaxy derives a SFR based on the integrated H$\alpha$ luminosity, and the axis ratio of the galaxy is taken from the NASA-Sloan Atlas catalog. Finally, given that broad line regions in AGN can cause overestimation of H$\alpha$-derived SFRs and/or mimick the presence of outflows in ionised gas tracers, we choose to identify and remove objects with AGN signatures in their central regions using a \citet{ka03} BPT diagnostic. Motivation for this also comes from several recent studies (e.g., \citealt{sarzi16,rb18,concas17}) which have demonstrated the limited influence of a weak AGN in the normal galaxy populations. Our resulting sample consists of 422 star-forming galaxies. To ensure our sample is not contaminated by the presence of mergers, stars, or pointing offsets, we visually inspect the MaNGA footprint image of each galaxy: 17 galaxies galaxies fail these criteria and are removed from the sample. Our final sample, therefore, consists of 405 star-forming galaxies, whose position on the SFR-M$_{*}$ plane is shown in Figure \ref{fig:sample} along with histograms of their SFRs$_{H\alpha}$ and stellar masses. The sample spans virtually the full MaNGA redshift range, with a median redshift of $z$=0.04 which corresponds to a pixel sampling of $\sim$0.4 kpc and an effective spatial resolution of FWHM$\sim$2 kpc.

\section{Stacking Procedures and Analysis}
\label{sec:stacking}
\subsection{Maps of Galaxy Properties}
As a first step in our analysis, we create maps of spatially-resolved galaxy properties (i.e., SFR, $\Sigma_{\text{SFR}}$, M$_{*}$, $\Sigma_{*}$, A$_{V}$ and D(4000)) using the MaNGA DR15 Pipe3D IFU maps for each galaxy in our selected sample. We begin by using the spatially resolved H$\alpha$ and H$\beta$ emission, for which the underlying continuum has been removed by the MaNGA DAP, in order to derive a Balmer decrement for each spaxel, which we translate into an A$_{V}$ and A$_{H\alpha}$ magnitude, assuming an intrinstic ratio of H$\alpha$/H$\beta$=2.68. The maps of H$\alpha$ are subsequently corrected for dust and converted to a luminosity using luminosity distances derived with an H$\alpha$ redshift and the assumed cosmology. The H$\alpha$ luminosities are subsequently converted to a $\Sigma_{\text{SFR}}$ using a \citet{kennicutt98} prescription converted to a Chabrier IMF (SFR [M$_{\odot}$yr$^{-1}$] = L$_{H\alpha}$/[2.1x10$^{41}$ erg/s]) and the physical area probed by each 0.5$''$ spaxel. Not all spaxels, however, are appropriate for analysis and we therefore ensure a quality control by applying the MaNGA bitmask flags and require the following criteria for science use:

\begin{itemize}
  \item A line S/N$>$3 for H$\alpha$, H$\beta$, [\OIII]$\lambda$5006, [\NII]$\lambda$6583 and [\OI]$\lambda$6300.
  \item A BPT ``star-forming'' nature determined by the combination of the [\NII]$\lambda$6583 \citet{ka03} and [\OI]$\lambda$6300 \citet{kewley06} prescriptions.
  \item An $r$-band S/N$>$5 to guard against spaxels with very little continuum signal.
\end{itemize}

\begin{figure*}
\center
 \includegraphics[width=0.7\textwidth]{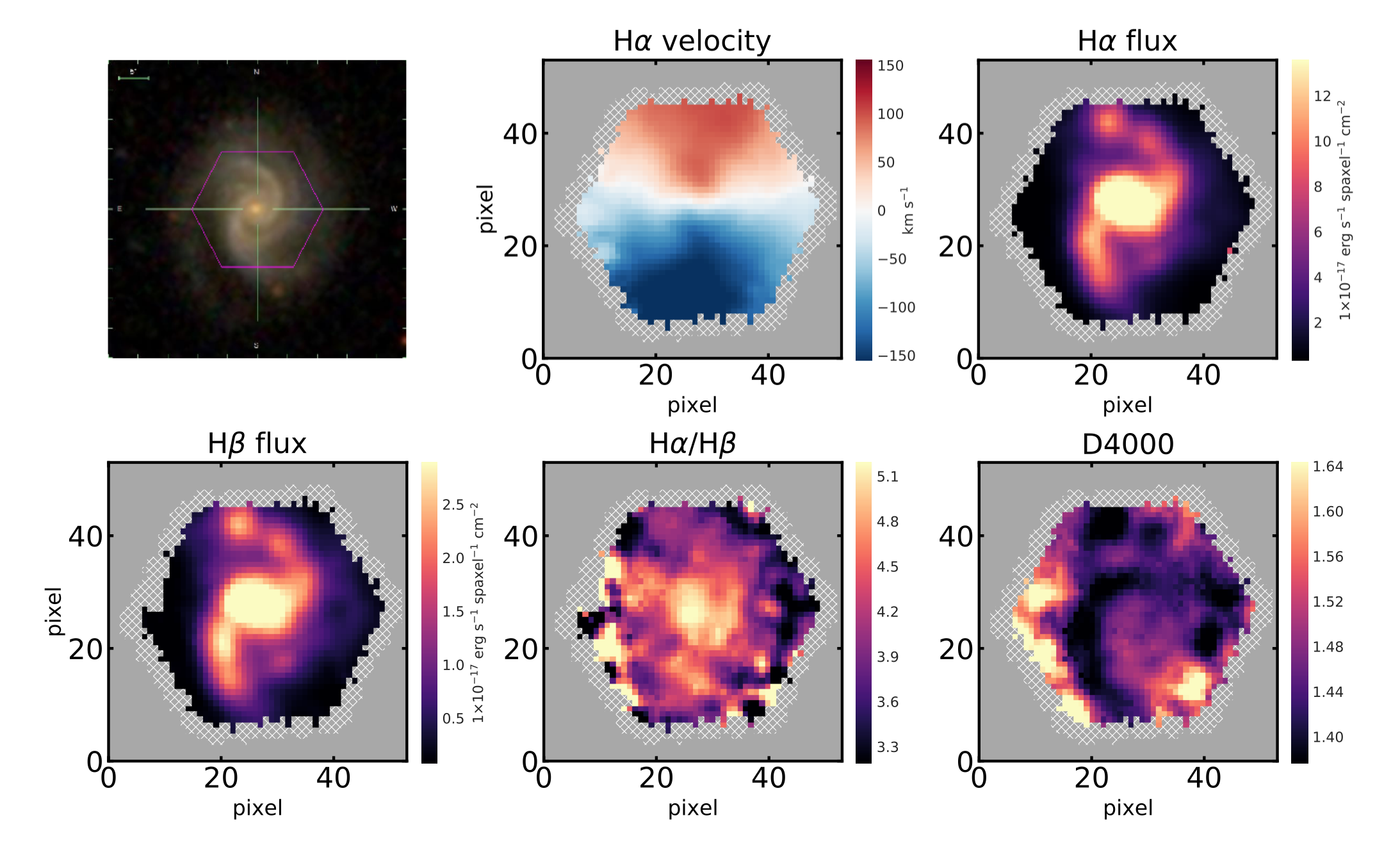}
 \caption{An example of IFU MaNGA galaxy property maps using the Marvin Python package. From left to right, top to bottom: The SDSS image of an example galaxy with the MaNGA footprint, the velocity of H$\alpha$ relative to the systemic, the flux of H$\alpha$, the flux of H$\beta$, the Balmer decrement, the D(4000) index.}
 \label{fig:maps}
\end{figure*}

The redshifts for each spaxel are derived from the \citet{talbot18} value added catalog\footnote{\url{https://www.sdss.org/dr15/data_access/value-added-catalogs/?vac_id=manga-spectroscopic-redshifts}}. 15 galaxies do not have determined spaxel redshifts and we exclude these from our analysis, resulting in a sample of 390 galaxies. In total, this results in 276,619 science spaxels and an example of the MaNGA maps for a representative galaxy in our sample is shown in Figure \ref{fig:maps}. We find that the planes of galaxy properties as traced by the full spaxel sample are fully sampled by each of the galaxy in our sample, ensuring stacks over any particular region of parameter space include virtually the full sample of galaxies. 

\subsection{Stacking Procedure and Outflow Modelling}
\label{subsec:proc}
Throughout this study we make use of stacking approaches to construct high S/N composite spectra with which to analyse outflow properties as a function of galactocentric radius and a variety of galaxy or spaxel properties.
The spaxels are first divided into a set of bins according to a given property and each spectrum in an associated bin is subsequently corrected for foreground galactic extinction using the associated \citet{schlegel98} E(B-V) values and an \citet{odonnell} Milky Way extinction curve, after converting the wavelength arrays of the spectra to air wavelengths. Each spectrum is then shifted to the rest-frame, before being interpolated over a common wavelength grid.

We subsequently mask all flux points in a given spectrum that are deemed unfit for science by its associated mask array and normalise the spectrum by the median flux between 5450\,\AA\ and 5550\,\AA\ (since this region is free of absorption and emission lines), thereby giving equal weight to each spectrum. The final spectrum is then taken as the mean over all the normalised spectra and the associated uncertainties are derived by adding in quadrature the bootstrapped sampling uncertainties and the mean flux uncertainties of each individual spectrum within the stack. In order to model outflow quantities associated with \nad, we use the same approach as described in \citet{rb18} to model the line and use the blueshifted absorption component to derive outflow quantities. We refer the reader to their study for details, however we summarise here the basic procedure of the approach: each \nad\ profile is first fit with an analytical model (where the intensity of \nad\ is based on a velocity-independent covering fraction, linewidth, column density and a possible velocity offset. See \citealt{rb18} for details) of a single fixed and single offset profile to determine whether a flow is present, and subsequently characterised with a three-component profile (consisting of a systemic component, blueshifted absorption and redshifted emission) if an outflow is deemed present. We use the same priors discussed in \citet{rb18}, however given the low covering fractions found in recent studies we limit the allowed values to $|\text{C}_{f}|\leqslant$0.5.

The mass outflow rate can be estimated through knowledge of the local covering fraction (or clumpiness of the gas), C$_{f}$, hydrogen column density, $N$(H), wind velocity, $v_{\text{out}}$, and an assumed geometry and radius. Whilst most absorption studies of outflows derive mass outflow rates based on a spherically symmetric thin shell geometry emanating from the centre of the galaxy, the resolved nature of the data sets use here means our stacks do not necessarily follow a well defined geometry. This is particularly relevant when stacking spaxels from different regions of a galaxy. Whilst one could derive rates based on the area probed instead, such an approach would result in direct correlation between the mass outflow rate and the number of spaxels in each stack, thereby removing much of any correlation with galaxy property. As such, we opt only to assume a radius of the outflowing gas of 1\,kpc - where the outflowing gas is unlikely to be collimated by the disk - and not to assume a geometry. The advantage of this is that our absolute values are less subject to uncertain geometrical assumptions and are almost entirely derived from our fitted parameters, whilst the disadvantage is that the true values are likely to be higher than what we present here. Thus, our expression for a mass outflow rate becomes:

\begin{equation}
\label{eq:newdmdt1}
\dot{M}_{\text{out}} = C_{f}\,\mu\,m_{\text{H}}\,N(\text{H})\,v\,r,
\end{equation}
where each term is as defined above, and m$_{\text{H}}$ is the mass of hydrogen, with an assumed correction for the abundance of Helium, $\mu$=1.4. Unless explicitly stated otherwise, we assume this general formulation for the mass outflow rate throughout all sections of this paper, and the assumption of $r$=1 kpc also remain unchanged throughout.

\section{The Galactocentric Profile of Outflows}
\label{sec:galradius}
\begin{figure}
\center
 \includegraphics[width=0.7\columnwidth]{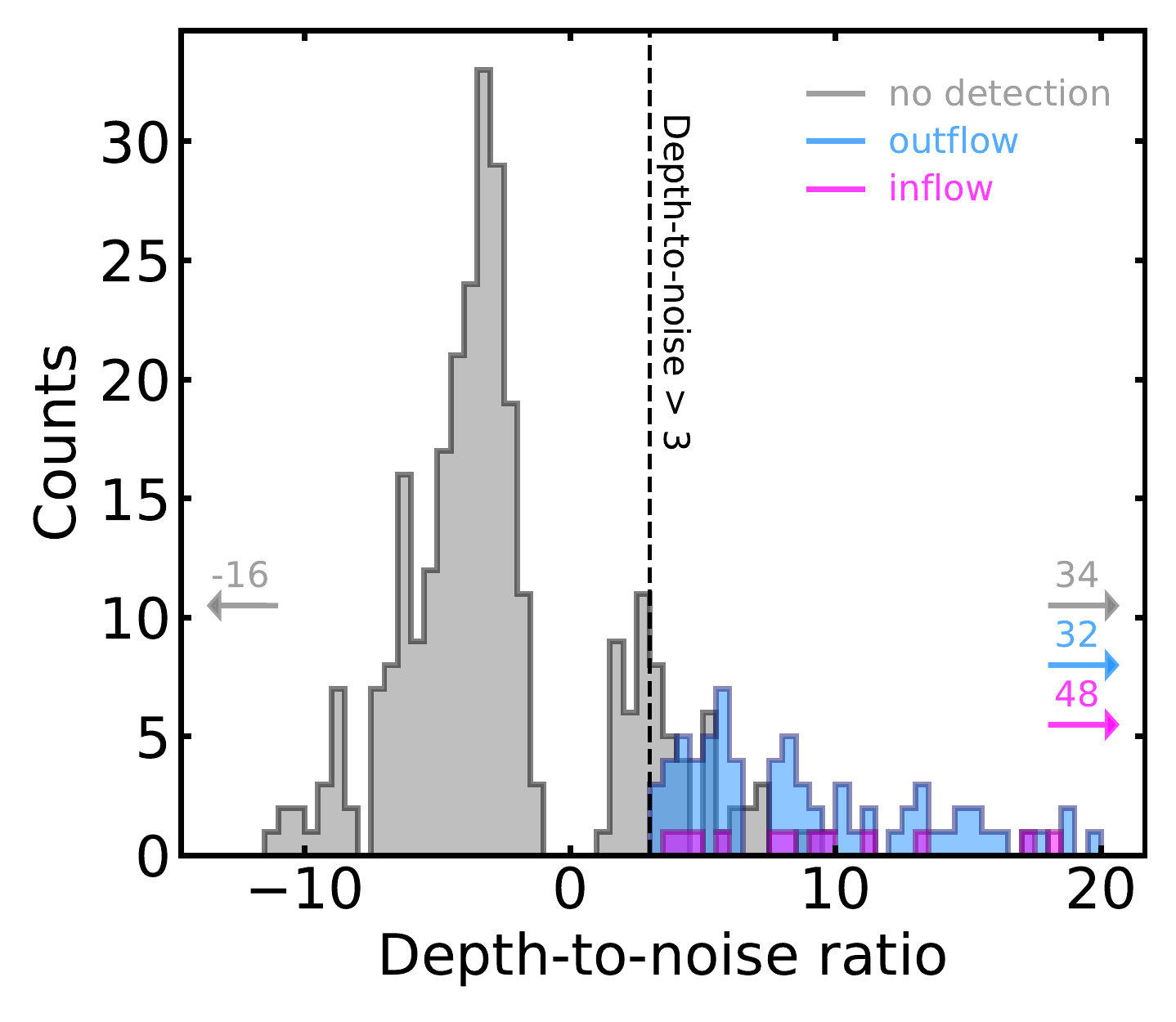}
  \caption{The typical depth-to-noise ratios of \nad\ in the central ($<$0.25\,R$_{e}$) regions of the sample of MaNGA galaxies used in this study. The gray histogram represents non-detections, whilst the blue histogram and the magenta histgram represent outflow and inflow detections, respectively. Positive values reflect profiles in absorption, whilst negative values represent profiles in emission. The dashed black line marks our minimum D/N threshold of $>$3 to determine detections of inflows and outflows.}
 \label{fig:dn_ratio}
\end{figure}

We begin by stacking spaxels as a function of deprojected galactocentric radius (i.e., accounting for the galaxy's inclination and rotation on the plane of the sky) for our sample of 390 galaxies, in order to create multiple independent annuli for each of our galaxies which probe the full range of the galaxy disk. The first annulus is centred at the centre of the galaxy disk and subsequent annuli extend outward from the border of the inner annulus. Each annulus contains spaxels within a full width of 0.25\,R$_{e}$ from its centre, and the distribution of annuli probe from the centre of the galaxy out to $\sim$2\,R$_{e}$. For each galaxy, all relevant spaxels are deprojected from the position of the central spaxel using the galaxy's position angle (PA) and inclination:

\small
\begin{equation}
\label{eq:int}
\text{RA}_{\text{deproj}} = (\text{RA}-\text{RA}_{\text{central}})\cdot\,\text{cos(PA)} + (\text{DEC}-\text{DEC}_{\text{central}})\cdot\,\text{sin(PA)}
\end{equation}

\begin{equation}
\text{DEC}_{\text{deproj}} = \frac{-(\text{RA}-\text{RA}_{\text{central}})\cdot\,\text{sin(PA)} + (\text{DEC}-\text{DEC}_{\text{central}})\cdot\,\text{cos(PA)}}{\text{cos}(i)}
\end{equation}
\normalsize
The deprojected radius between a given spaxel and the central spaxel is then taken to be r$_{\text{deproj}}$=$\sqrt{\text{RA}^{2}_{\text{deproj}} + \text{DEC}^{2}_{\text{deproj}}}$, and the mean spectrum and quantities quoted here are those over the central spaxels. The motivation here is to determine and select galaxies that display signatures of outflows, in order for them to be used in subsequent analyses where we can tie global galaxy properties to outflow properties. Since the signature of blueshifted absorption can easily be ``diluted'' by strong systemic absorption in a stack, we wish to maximise our chances of outflow detection and characterisation by preselecting galaxies with outflows in their central regions. Thus, to determine whether a gas flow is present in the central region (R$<$0.25\,R$_{e}$), we compare \nad\ fits of a fixed systemic component to a single blueshifted or redshifted absorption component. For this we use a Bayesian Information Criterion (BIC), which penalises for extra free parameters and favours lower values. We derive ratios (K=BIC$_{\text{fixed}}$/BIC$_{\text{flow}}$) of the two models to account for the extra free parameter (i.e., a velocity offset) in the shifted model and determine a flow detection if the ratio is greater than unity and a minimum blueshift velocity is found. Furthermore, several recent studies have shown that Sodium excess can sometimes be seen in emission \citep{chen10,rb18,concas17}, with the rate of outflow detections decreasing rapdily as the net profile changes towards emission. Given that the nature of the \nad\ emission is still poorly understood, we opt to analyse only galaxies which show net absorption at their centres, and require an absorption depth-to-noise (D/N) ratio $>$3 in the central region, with a minimum blue shifted of 15 km\,s$^{-1}$ if D/N$>$5 and 20 km\,s$^{-1}$ if 3$<$D/N$<$5. In total, 92 galaxies satisfy these criteria, with 78 objects displaying outflow detections and 14 displaying inflow detections. The remaining 298 galaxies either do not show sufficient absorption ($\sim$80\%), with a typical (median) D/N ratio of 1.62, do not show blueshifted absorption ($\sim$18\%), or have insufficient spatial resolution to stack within 0.25\,R$_{e}$ ($\sim$2\%). 
We report the flow galaxies and their main properties in Table \ref{tab:samplegals} and illustrate the typical D/N ratios in Figure \ref{fig:dn_ratio}.

\subsection{Average Central Profiles}
\begin{figure*}
\center
 \includegraphics[width=\textwidth]{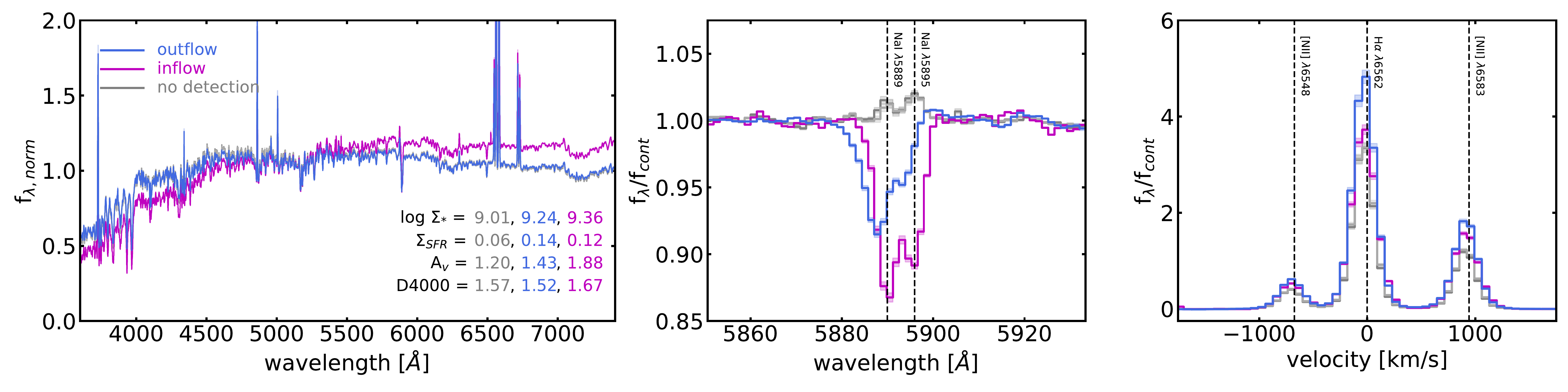}
 \caption{The average central 0.25\,R$_{e}$ spectra of galaxies harbouring outflows (blue) or inflows (magenta), and the average spectra of an associated control sample of non-detection galaxies (light gray for inflows, dark gray for outflows). The 1$\sigma$ scatter is shown as the colour shaded regions around each spectrum. \textit{Left:} the normalised optical spectra over virtually the full optical range. \textit{Middle:} the continuum-normalised ISM residual of the \nad\ line for the four spectra in the left panel. \textit{Right:} the continuum-normalised H$\alpha$+[\NII] emission for the four spectra in the left panel. The quantities and units quoted in the left panel are the stellar mass surface density (M$_{\odot}$kpc$^{-2}$), SFR surface density (M$_{\odot}$yr$^{-1}$kpc$^{-2}$), dust (magnitudes) and 4000\,\AA-break. The colours correspond to the different stacks (NB: given that the average spectra - and quantities - for the two non-detection control samples are essentially identical, we display only the values of the control sample associated with the outflow detections).}
 \label{fig:meanspecs}
\end{figure*}

We begin by examining and comparing the mean spectra of galaxies with and without flow detections in the inner 0.25\,R$_{e}$ of their effective radii. We adopt a Monte Carlo approach to constructing the mean central stacks over samples of outflow and inflow galaxies, using a control sample of non-detection galaxies for comparison: a random sample of 10 outflow (or inflow) galaxies is selected and for each galaxy we randomly select a non-detection ``counterpart", defined as such if it lies within $\pm$0.2 dex in stellar mass and SFR, before averaging the central spectra. This is repeated 100 times and the final spectrum is simply the mean over all iterations, with the stack flux errors derived from a combination of individual flux uncertainties and bootstrapping errors. The results of these are shown in Figure \ref{fig:meanspecs}, where we show the differences in spectra between galaxies with outflows, inflows, and those without, along with some key galaxy properties. From the left panel, we note that the optical spectra over $\sim$3700$<\lambda<$8000\,\AA\ for outflow and non-detection galaxies are virtually identical, with minor differences only in the strength of the emission lines and depth of the absorption lines. The similarity is less pronounced for inflow galaxies, however, which display a considerably redder continuum slope than their non-detection counterparts. To quantify these differences, we take the mean continuum fluxes between 4500 $\leqslant\lambda\leqslant$ 4800 \AA\ and  5500 $\leqslant\lambda\leqslant$ 5800 \AA\ - roughly equivalent to the $B$ and $V$ bands - and find $B-V$ colours of 0.015$\pm$0.039 mag and 0.003$\pm$0.041 mag for outflow galaxies and their associated non-detections, respectively, and 0.132$\pm$0.045 mag and 0.001$\pm$0.041 mag for inflow galaxies and their associated non-detections, respectively. In the case of outflow galaxies and both non-detection spectra, the spectra display virtually no $B-V$ excess, suggestive of a flat spectrum. However, this contrasts with the spectra of inflow galaxies which display a more significant $B-V$ excess ($\sim$0.12 mag redder than the outflow galaxies), and fitting straight lines to the pseudo-photometry reveals an inflow galaxy spectral slope $\sim$9$\times$ greater than that of the outflow spectra.


A much starker contrast, however, appears when comparing the average \nad\ profiles (middle panel) profiles. For \nad\, we note a large difference in the total equivalent width (EW) of the profile, with $\sim$0.5\,\AA\ for outflow detections, $\sim$0.95 for inflow detections, and -0.1\,\AA\ for non-detections. A visual comparison of the profiles shows that both inflow and outflow detections are characterised by significant and unambiguous absorption. The \nai $\lambda$5889 line has an intrinsic line depth twice the size of its \nai $\lambda$5895 redshifted counterpart \citep{morton91}, however the flow profiles display a clear asymmetry: the outflow spectrum is blueshifted with respect to the systemic wavelengths and the inflow spectrum, while not as significantly redshifted, shows nearly equal line depth for the two \nad\ lines which can arise due to an additional redshifted component. Both profiles are well-characterised by an additional blueshifted or redshifted component offset by $|\Delta$v$|\sim$120 \kms\ and $|\Delta$v$|\sim$85 \kms, respectively. The average non-detection spectra, on the other hand, are characterised by virtually no absorption and even display net emission. In stacking the full non-detection sample, we find this emission becomes even more pronounced. 

Although less pronounced, similar comparisons are found with the ionised gas traced by H$\alpha$+[\NII] (right panel): the average non-detection spectrum is characterised by a narrow profile that is less luminous than its detection counterpart, by a factor of $\sim$0.7. In contrast to the outflow observed in the mean \nad\ profile, a significant broadening of the ionised gas is not immediately obvious, however a BIC ratio between a one- and two-component fit strongly favours the latter fit, with an outflow velocity of $\sim$130 \kms, similar to that observed with the neutral gas and suggestive of the emission emanating both from the disk of the galaxy and a broader outflowing component.

To gain some indication of the driver for the different profiles, we compare the mean values of several galaxy properties likely to be important in determining both the shape of the \nad\ and H$\alpha$+[\NII] profiles and flow presence. \nad\ requires significant amount of shielding from dust to survive, and dust obscuration has been found to correlate with the EW of the line (e.g., \citealt{veilleux95,heckman2000,chen10, rupke13,rb18}). Here we find average A$_{\text{v}}$ magnitudes of $\sim$1.4-1.9 mag (detections) and $\sim$1.2 mag (non-detections), suggesting at first glance somewhat of a limited impact in regulating the shape of the two flow profiles. Such a conclusion may come as a surprise given previous results in the literature, however several reasons exist to explain this. The first is simply that dust is not the only regulating property of the EW of \nad, since inclination and gas column densities will also impact the shape of the profile: indeed, \citet{chen10} find a strong correlation between NaD EW and inclination for SDSS galaxies which probe similar nuclear areas to our samples of MaNGA galaxies. With inclinations lower than i$\sim$40$^{\circ}$, they observe systemic \nad\ transitioning from absorption into pure emission, possibly due to smaller column densities of gas probed along the line of sight through the thickness of the disk and difficulties of the continuum-fitting code to reproduce such low column densities. If we compare our results to theirs, we note that the median inclination of our high mass MaNGA galaxies displaying net emission is significantly lower at $\sim$35$^{\circ}$, enough to push the \nad\ profile and EW into net emission. Finally, we do point out that even over the small range of mean A$_{V}$ values, the relation between \nad\ EW and A$_{V}$ is still present when comparing the values to the size of the \nad\ profiles in the middle panel of Figure \ref{fig:meanspecs}.

Similar observations are found for a dependence on the D(4000) break, which varies only by $\sim$0.1 between the spectra, suggesting limited impact by the age of the underlying stellar population. The difference in $\Sigma_{\text{SFR}}$, on the other hand, we find to be much more significant: for non-detections, the average value (0.048$\pm$0.001\,M$_{\odot}$yr$^{-1}$kpc$^{-2}$) is lower compared to the average values of detections, which are both higher than $>$0.1\,M$_{\odot}$yr$^{-1}$kpc$^{-2}$ (0.116$\pm$0.001 and 0.126$\pm$0.002 for outflows and inflows, respectively). Although the differences between the average values is not, at first glance, particularly large or significant, it is their absolute values which is of greater interest: the $\Sigma_{\text{SFR}}$s of the flow spectra are consistent with (and above) what is generally invoked as a minimum threshold for outflow activity seen in absorption (0.1\,M$_{\odot}$yr$^{-1}$kpc$^{-2}$; \citealt{heckman02}). Finally, we also find elevated stellar mass surface densities for the detection spectra, compared to the non-detections and this is most likely the result of a positive correlation with $\Sigma_{\text{SFR}}$ (i.e., the ``resolved'' Main Sequence; e.g., \citealt{sanchez13,canodiaz16,gonzalez16,abdurro17,hsieh17,maragkoudakis17}). We attempt to disentangle these intrinsic correlations and explore further the main regulating properties of the outflows in Section \ref{sec:galprops}.

\subsection{The Radial Extent of Outflows and Their Properties}
\label{subsec:radial}
\begin{figure*}
\center
 \includegraphics[width=\textwidth]{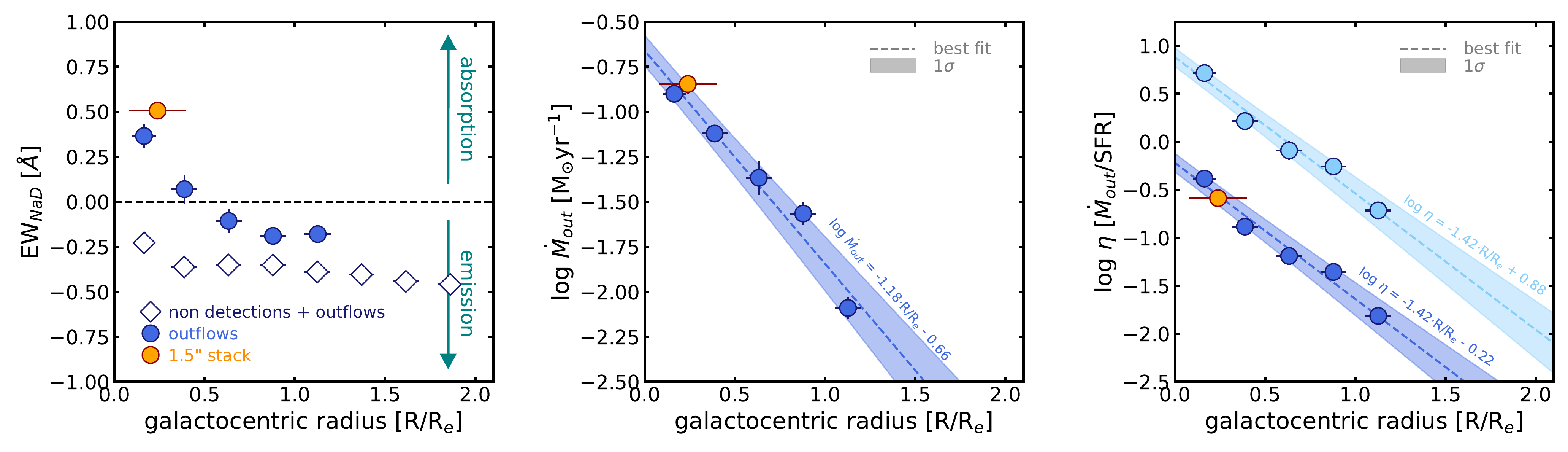}
 \caption{The evolution of the total \nad\ EW (left panel), mass outflow rate (middle panel) and mass loading factor (right panel) as a function of galactocentric radius. 1$\sigma$ error bars are included in each panel and in many cases are smaller than the symbols themselves. The \nad\ EW is separated out into stacks of spaxels from the outflow detection sample only (circles) and non-detection+outflow spaxels (diamonds). Filled symbols indicate the detection of outflowing neutral gas, whilst empty symbols indicate non-detections. For comparison, a 1.5$''$ stack is also added to each of the plots, in order to gauge what would be measured by single-fiber SDSS surveys. In the middle and right plots, we add a best fit first order polynomial (dashed line) and its 1$\sigma$ error (shaded region). To illustrate the impact of different geometric assumptions in the calculation of mass outflow rates and loading factors, in the right panel we plot the mass loading factor assuming no outflow geometry (dark blue points and orange SDSS point; as described in Section \ref{sec:stacking} and by Equation \ref{eq:newdmdt1}) and a spherically symmetric thin shell geometry of 4$\pi$ (light blue points; these are simply the dark blue points multiplied by 4$\pi$, see Section \ref{subsec:radial} for details). These serve as lower and upper limits of the mass loading factor, respectively.}
 \label{fig:radial}
\end{figure*}

Outflows in star-forming galaxies are typically thought to display a degree of collimation in their geometry, most likely attributed to the underlying disk pressure gradient resulting from the interplay between gravity and the outward momentum from stellar feedback. Here, we test this picture by looking at the extent and properties of outflows over a range of galactocentric radii, R/R$_{e}$. This is shown in Figure \ref{fig:radial}, where we plot the evolution of the ISM \nad\ EW (left panel), mass outflow rate (middle panel), and mass loading factor (right panel) as a function of radius. For reference, we also compare these to what would be measured by an SDSS 1.5$''$-radius fiber.

In the left panel of Figure \ref{fig:radial} we plot the EW of the \nad\ residual - measured directly from the spectrum itself - for stacks containing only the outflow galaxies (circles) and stacks with both the non-detection and outflows galaxies (diamonds). For the former, we observe a rapid decrease in EW as a function of radius, with a roughly exponentially declining profile which begins to flatten out slightly at $>$0.75\,R$_{e}$. The central regions of the galaxy ($<$0.5\,R$_{e}$) are dominated by absorption, although this quickly transitions to net emission outward of $\sim$0.5\,R$_{e}$. A 1.5$''$ stack of the detection galaxies shows that SDSS measurements would be consistent and similar to what we measure here. Although we cannot compare directly, we note that this is similar to the distribution of molecular gas (i.e., cold gas) in disk galaxies traced by CO \citep{bigiel08,leroy08,schruba11,bigiel12}, where observations of molecular gas as a function of galactocentric radius have found a roughly exponential decrease. Outflow detections are observed only out to $\sim$1\,R$_{e}$, with the majority of the blueshifted absorption occurring within 0.5\,R$_{e}$ and the outer regions detected primarily due to the imprint of redshifted emission that backscatters off of the receding side outflow. We note that some degree of bias may exist in these observations, since the detection of blueshifted absorption is contingent upon observing \nad\ in absorption, which in turn is partially regulated by the strength of the underlying continuum. Since this latter consideration is driven by the surface density of stars, which drops off steeply at higher radii, non-detections of outflows due to the lack of \nad\ absorption are likely to occur in regions of low $\Sigma_{M_{*}}$. As such, it is possible our detections are biased towards the central regions of the galaxies where higher stellar mass surface densities are found, however our choice to stack in units of R/R$_{e}$ should provide some safety against such a bias. 

A milder trend is observed in stacks containing both the non-detection galaxies and outflow galaxies (diamond symbols in the left panel of Figure \ref{fig:radial}), albeit with some important differences: the strength of \nad\ emission increases slightly with radius, but in general maintains a much flatter slope compared to when only outflow galaxies are considered. Additionally, the added presence of the non-detection galaxies ensures the EW never reaches absorption at any radius and is observed only in emission; no detections of outflows are observed in these stacks. The contrast between the two samples is particularly evident below $\sim$1\,R$_{e}$, where outflow detections are present, and becomes less great outward of this limit. Such a comparison highlights the ease with which signals of outflows in normal galaxies can be overlooked if not selected appropriately. It is unclear, however, what induces such a rapid change in the \nad\ profile. Although typically assumed to be an absorption transition, several studies using SDSS \citep{chen10,rb18,concas17} have also found \nad\ in emission in stacked spectra. One obvious hypothesis is that SSP models at such low \nad\ EW are unable to accurately reproduce the stellar contributions and as such overestimate them, although there is a possibility that at outer radii there is insufficient gas and dust to produce significant \nad\ absorption and potential background \nad\ re-emission begins to dominate. 

We also wish to determine the radial evolution of the mass outflow rates in our stacks. To do this, we use Equation \ref{eq:newdmdt1} to derive values of $\dot{M}_{\text{out}}$ for the detections presented in the left panel of Figure \ref{fig:radial} and plot these as a function of radius in the middle panel. In these stacks, the adopted $N$(H) value is the mean value of the summed column density for a given annulus, over all galaxies in the stack. We immediately note an important decrease in $\dot{M}_{\text{out}}$ as a function of galaxy radius, similar to the trend observed for the \nad\ EW in the left panel. The outflow appears strongest in the central regions of the galaxy, with values starting at $\sim$0.13\,M$_{\odot}$yr$^{-1}$ and decreasing down to $\sim$0.01\,M$_{\odot}$yr$^{-1}$. 
A comparison with the 1.5$''$ stack shows SDSS observations only probe the strongest parts of the outflow, with a mass outflow rate of 0.14\,M$_{\odot}$yr$^{-1}$, and likely overlook important contributions from the outer regions (i.e., out to $\sim$1\,R$_{e}$). Specifically, an SDSS 3$''$ fiber would miss $\sim$50\% of the total mass outflow rate integrated over all the MaNGA radial stacks, which we calculate to be $\dot{M}_{\text{out}}\sim$0.28\,M$_{\odot}$yr$^{-1}$. The trend observed by the MaNGA points is well described by a linear fit, with a slope of $\sim$-1:

\begin{equation}
\text{log}_{10} \dot{M}_{\text{out}} = (-1.18\pm0.12)\cdot R/R_{e} - (0.66\pm0.08) \\
\end{equation}

Such a trend is perhaps not surprising: the outflows are selected to be star formation-driven and as such are likely to correlate with star formation- and gas-dependent quantities, which are known to be mostly centrally concentrated (e.g., \citealt{bigiel08,ellison18}).

The derivation of mass outflow rates in conjunction with the mean SFRs associated with each ring stack allow us to compute a resolved mass loading factor, $\eta$, and to first order, determine the extent of the outflows' potential for quenching. The mass loading factors are shown in the right panel of Figure \ref{fig:radial}. We observe a very similar trend to the mass outflow rate, with a rapidly decreasing log\,$\eta$ as a function of radius, characterised by a slightly steeper slope of -1.4.

\begin{equation}
\text{log}_{10} \eta = (-1.42\pm0.15)\cdot R/R_{e} - (0.22\pm0.10)
\end{equation}

The values range from mass loading factors of $\eta\sim$0.4 in the central regions of the outflow and decrease to $\eta\sim$0.02 in the outermost regions, suggesting the potential for ejective quenching by the outflows is strongest in the central regions and not a galaxy-wide phenomenon - i.e., only in the central regions does the outflow have any sort of potential to quench the galaxy host by removing gas faster than the rate of star formation and halting star formation activity - although none of the values ever reach unity, suggesting quenching even in the central regions remains unlikely. Integrating over all outflow detections, we find the galaxy-wide mass loading factor is $\eta\approx$0.1, supporting this hypothesis.

However, as mentioned in Section \ref{subsec:proc}, our choice not to assume an outflow geometry for these outflow rates means we are, to a degree, likely underestimating the absolute values. As such, for comparison we calculate mass loading factors with outflow rates derived with an assumed geometry. We assume here the outflowing gas from each stack is coming from an individual ``wind bubble'', which we measure as a spherically symmetric shell subtending 4$\pi$ steradians, whose origin resides in the galaxy disk and extends out to 1\,kpc. The base assumption here is that the large-scale outflow is formed via the superposition and collimation of such ``wind bubbles'' at larger radii (e.g., 5\,kpc) above the disk. Thus, we multiply the outflow rates derived with Equation \ref{eq:newdmdt1} by 4$\pi$ and show these as light blue points in Figure \ref{fig:radial}. The result shows mass loading factors $\sim$1.1 dex greater than our fiducial measurements, with a range 0.2$\lesssim \eta \lesssim$5.2, suggesting an enhanced possibility of first order ejective quenching in the central regions of the galaxy and highlighting the importance of geometrical assumptions. We note that for a spherically symmetric thin shell geometry at 1\,kpc, the two ranges of outflow rates presented here reflect the lower and upper limits of the model.

\begin{figure}
\center
 \includegraphics[width=0.85\columnwidth]{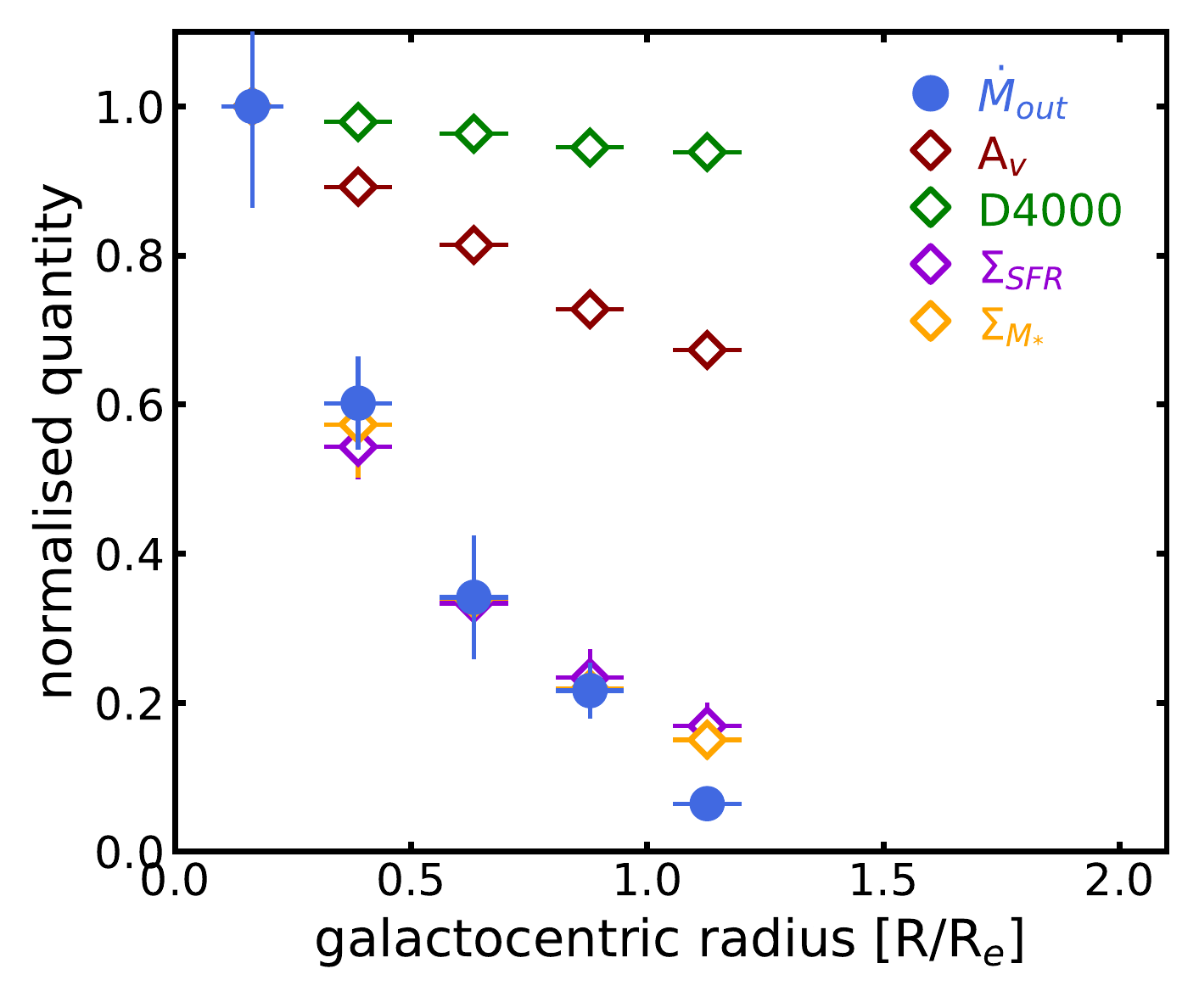}
 \caption{The normalised evolution of the geometry-independent mass outflow rate compared to several key galaxy properties (i.e., $\Sigma_{\text{SFR}}$, $\Sigma_{M_{*}}$, A$_{V}$, D(4000)), with error bars. Each quantity is normalised by the maximum value across galaxy radius and error bars of galaxy properties include - for each annulus - the standard deviation over all spaxels and all galaxies going into the stack. The radial evolution of the normalised mass outflow rate is most closely aligned with the normalised evolution of $\Sigma_{\text{SFR}}$ and $\Sigma_{M_{*}}$.}
 \label{fig:dmdt_evol}
\end{figure}

It is unclear whether it is a radial dependence or the average galaxy properties which drive the main trends seen in Figure \ref{fig:radial}. In fact, by comparing galaxy properties associated with each radial bin to the evolution of the mass outflow rate, we find the latter quantity follows most closely the evolution of $\Sigma_{\text{SFR}}$ and $\Sigma_{*}$. This is shown in Figure \ref{fig:dmdt_evol}, where we plot the normalised evolution of galaxy properties and the mass outflow rate. However, from these stacks alone it is difficult to determine whether the selected properties or the radial dependence are the primary regulators of the outflows, given that intrinsic correlations exist between galaxy radius and the chosen properties. We further inspect the correlation of (or lack of) the \nad\ EW and outflow properties with global galaxy properties in Section \ref{sec:galprops}.

\section{The Resolved $\Sigma_{\text{SFR}}$-$\Sigma_{M_{*}}$ Plane}
\label{sec:mainseq}
\begin{figure*}
\center
 \includegraphics[width=\textwidth]{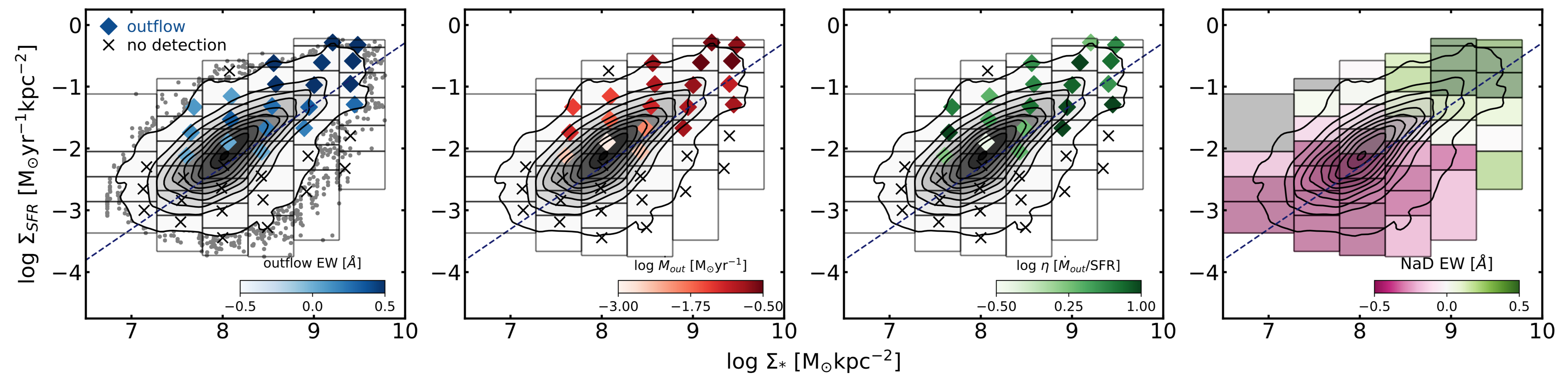}
 \caption{The resolved $\Sigma_{\text{SFR}}$-$\Sigma_{\text{*}}$ plane of star-forming MaNGA galaxies with stacked bin limits (gray lines) and outflow detections (diamonds) overplotted. Non-detections are marked by a cross and detections are colour-coded according to the mean outflow EW (left panel), neutral gas mass outflow rate (middle left panel), and local mass loading factor (middle right panel). Outflows are prevalent in regions of high star formation activity and stellar mass surface density, with an EW and mass outflow rate that follows a similar pattern. For the mass outflow rate and loading factor, no outflow geometry is assumed. The right panel is the same as the other panels except colour-coded by the total \nad\ EW, as measured from the stacked spectrum. The dashed navy line indicates the resolved MS relation for \Hii\ regions of star-forming galaxies from \citet{hsieh17}.}
 \label{fig:sfrmstarplane}
\end{figure*}

Since the discovery of the galaxy Main Sequence (MS) and the development of basic frameworks to describe a galaxy's position relative to it, much work has gone into determining the prevalence and influence of outflows relative to the MS (e.g., \citealt{chen10,rubin14,cicone16,rb18,concas17}). The arrival of large IFU surveys such as MaNGA, SAMI and CALIFA, has also revealed a resolved MS \citep{sanchez13,canodiaz16,gonzalez16,abdurro17,hsieh17,maragkoudakis17}, indicative of a link between small scale processes and the integrated properties of galaxies. As such, it is interesting to look at the prevalence and properties of outflows over such a resolved sequence. We therefore present a stacked analysis over the local MS in Figure \ref{fig:sfrmstarplane}, using spaxels from our sample of outflow galaxies. Here, the mean stacks and properties are taken over all spaxels in a bin, rather than over the galaxies going into the stack. From the left panel of Figure \ref{fig:sfrmstarplane}, we find a very similar trend of outflow prevalence and strength to what is found in integrated analyses: outflows are found predominantly in regions of high $\Sigma_{\text{SFR}}$ and $\Sigma_{M_{*}}$, with increasing strength (as traced by the outflow EW and mass outflow rates) higher up the MS. Here we find that detections span a $\Sigma_{*}$ range of 7.5$<$log $\Sigma_{*}$/M$_{\odot}$kpc$^{-2}<$9.5 and $\Sigma_{\text{SFR}}$ of -2$<$log $\Sigma_{\text{SFR}}$/M$_{\odot}$yr$^{-1}$kpc$^{-2}<$0, in agreement with limits found by \citet{rb18} for neutral gas outflows of similar galaxies.

Additionally, we also investigate the evolution of the outflow EW, (geometry-independent) mass loss rate and mass loading factor (shown by the colour-coding of detections in the first three panels of Figure \ref{fig:sfrmstarplane}). Unlike in Section \ref{subsec:radial}, the adopted $N$(H) value here is the mean column density \textit{per pixel}. We find the first two quantities correlate positively with the outflow's position along the resolved MS: spaxels higher up the MS produce stronger outflows with significantly larger EWs and mass loss rates, consistent with the notion that more intense star formation activity drives stronger outflows. The mass outflow rates range from -3$\lesssim$log $\dot{M}_{\text{out}}$/M$_{\odot}$yr$^{-1}\lesssim$0 with \HI\ column densities 18.6$<$log $N$(H)/cm$^{-2}<$22.0 and median values of log $\dot{M}_{\text{out}}\sim$-1.0 M$_{\odot}$yr$^{-1}$ and log $N$(H)$\sim$20.32 cm$^{-2}$, respectively. However, the picture is not as clear cut for the mass loading factor, which does not appear to display significant evolution as a function of MS position (although there is tentative indication of higher factors further up the MS). The values here range from -0.4$\lesssim$log $\eta\lesssim$1.5, with a median mass loading factor of log $\eta\sim$0.6.

From the right panel of Figure \ref{fig:sfrmstarplane}, we find that outflows detections are found predominantly in absorption and the lower limit of their detections corresponds to regions of reduced ISM absorption (or P-Cygni profiles), where the \nad\ profile transitions into emission. Profiles of pure emission generally do not display any outflow signatures. The total \nad\ profile also shows a similar transition across the plane as observed in the SFR-M$_{*}$ plane with 3$''$ fibers \citep{rb18,concas17}: absorption at high stellar mass and SFRs transits to emission at low values of the same quantities. However, the dichotomy of the two profile types with log $\Sigma_{\text{SFR}}$ is more evident than with log $\Sigma_{\text{*}}$, log SFR, or log M$_{*}$, and as such likely assigns $\Sigma_{\text{SFR}}$ as the main regulator of the ISM profile.

It is important to note that the evolution in outflow prevalence and properties seen over the local $\Sigma_{\text{SFR}}$-$\Sigma_{\text{*}}$ plane also corresponds to the same radial evolutions seen in Figure \ref{fig:dmdt_evol}, given that the evolution of these particular galaxy properties also evolve with galaxy radius (i.e., both $\Sigma_{\text{SFR}}$ and $\Sigma_{\text{*}}$ decrease with increasing radius). Thus, the strongest outflow detections are found predominantly in the central regions of galaxies.

\section{Outflow Correlations with Galaxy Properties}
\label{sec:galprops}
As hinted at by Figure \ref{fig:dmdt_evol}, the evolution of outflow properties is likely tied to one or several underlying galaxy properties. This has often been investigated for integrated galaxy quantities (such as total SFR, stellar mass and $\Sigma_{\text{SFR}}$; \citealt{chen10}), however such investigations generally rely on single or stacked spectra of the central regions of a galaxy, making the isolation of the regions of interest challenging.

\begin{landscape}
\begin{figure}
\centering
 \includegraphics[width=1.25\textwidth]{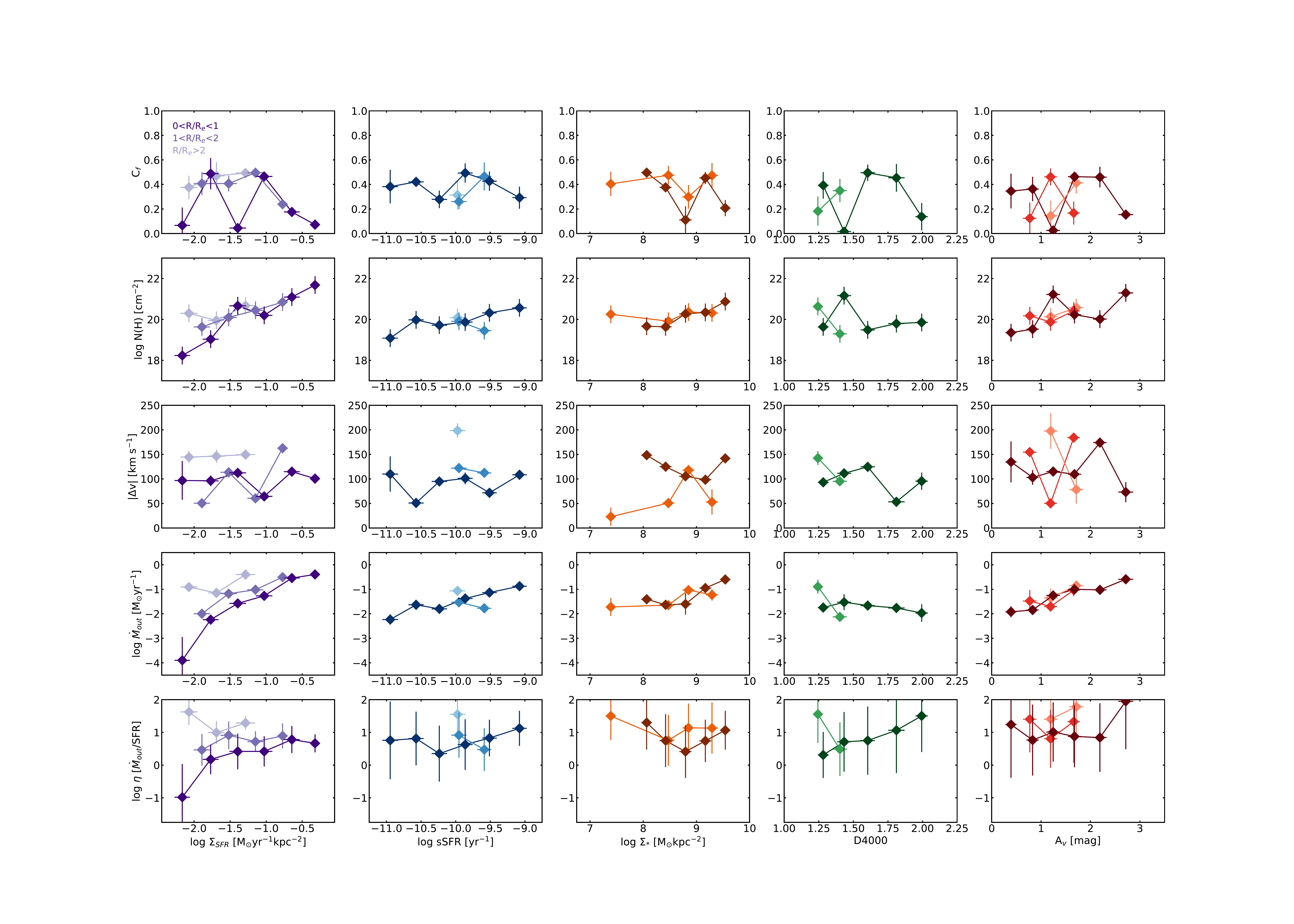}
 \caption{The correlations between outflow properties ($C_{f}$, $N$(H), $\Delta$v, $\dot{M}_{\text{out}}$ and $\eta$) as a function of galaxy properties as shown by stacks of individual spaxels within a given property range. The mass outflow rates and mass loading factors here are independent of outflow geometry. The main trends occur with star-forming quantities, although the strongest correlation is seen with $\Sigma_{\text{SFR}}$.}
 \label{fig:outflow_props}
\end{figure}
\end{landscape}

However, due to the power of IFU spectroscopy, we are in the fortunate position of being able to identify and separate individual spaxels (and therefore kpc-scale regions) in individual galaxies corresponding to a specific range of a given global galaxy property, thereby removing many of the intrinsic correlations that exist between a given property and e.g., galaxy radius. We therefore perform this analysis for key galaxy quantities such as $\Sigma_{\text{SFR}}$, $\Sigma_{\text{*}}$, specific SFR (sSFR), A$_{V}$, and D(4000), which for each galaxy we divide into galactocentric radial bins in order to eliminate underlying correlations between properties, and stack over all outflow galaxies. We present this analysis in Figure \ref{fig:outflow_props}. Our choice of galaxy properties is motivated in part due to the availability of the tracers from optical spectra, but more importantly due to their inferred influence on outflows from integrated studies: star formation-related quantities are invoked as the main drivers for outflows in the absence of an AGN (e.g., \citealt{heckman2000,veilleux05,chen10}), whilst both stellar mass and dust are influential in regulating the escaping potential of outflows and the survivability of \nad\, respectively. The D(4000) break traces the age of the underlying stellar populations, and as such can provide a first order indication as to which stellar populations may be driving the outflowing gas.

We begin by noting that outflows are detected over a large range of each galaxy property. The ranges span -2.25$\lesssim$log $\Sigma_{\text{SFR}}$/M$_{\odot}$yr$^{-1}$kpc$^{-2}\lesssim$-0.25, -11$\lesssim$log sSFR/yr$\lesssim$-9, 7.5$\lesssim$log $\Sigma_{\text{*}}$/M$_{\odot}$kpc$^{-2}\lesssim$9.5,
1.2$\lesssim$D(4000)$\lesssim$2.0, and 0$\lesssim$A$_{V}$/mag$\lesssim$3, consistent with ranges presented in all of our stacked analyses so far. Most of the detections are found in the inner radii, with the detection rate falling rapidly as a function of radius.


Next, we inspect the local covering fraction of the gas, C$_{f}$, which is an estimate of its clumpiness; low values indicate that the gas is clumpy and therefore not covering much of the background continuum source, whilst values closer to unity indicate a more diffuse component. For SDSS-selected galaxies, small values ($C_{f}<$0.5) for \nad\ gas have been found, and this is attributed to the low ionisation potential (5.1 eV) of the atom, which requires significant self- and dust-shielding to survive. As such, a more compact and clumpy nature ensures it can survive harsher environments (e.g., from shocks and hard stellar radiation fields). We find significant scatter in the covering fraction of the gas spanning the full range of allowed values, with no clear evolution as a function of galaxy property or galaxy radius. Given that these stacks are constructed over a variety of radii and the estimates of C$_{f}$ are velocity-independent measurements, the scatter is perhaps not surprising.

We also look at the column density of the outflowing hydrogen gas, since it is interesting to determine where the densest regions of the outflows reside. We find clear, positive correlations of the column density with log $\Sigma_{\text{SFR}}$ and log sSFR, and a minor correlation found with log $\Sigma_{\text{*}}$. No correlation is found with the D(4000) break and only a tentative trend with A$_{V}$ may be present in the inner radii of galaxies. The first three quantities all strongly correlate with the amount of cold gas present and the star formation activity of galaxy regions, therefore such trends are perhaps not surprising: higher concentrations of gas induce more star formation which produce stronger outflows which expel more gas. This is particularly evident from the strongest correlation with log $\Sigma_{\text{SFR}}$ which spans 4 dex in column density and suggests the densest parts of the outflowing gas correlate directly with the densest regions of star formation and cold gas.

In measuring the blueshift of absorbed gas, we report absolute outflow velocities of $\sim$50-200 \kms, similar to those reported for normal galaxies at $z\approx$0 \citep{chen10,sugahara17,rb18,rodriguezdelpino19}, however we observe essentially no correlations of the outflow velocity with any of the chosen galaxy properties. The evolution of the outflow velocity with galaxy properties and redshift has long been debated, with some studies claiming a velocity evolution over SFR \citep{sugahara17} and others showing little to no evolution \citep{martin05,chen10,rb18,rodriguezdelpino19}. Such debate is also subject to the manner of measuring outflow velocities: for instance, in stacked spectra at $z\sim$0, \citet{sugahara17} find convincing evidence for a velocity evolution across SFR when measured at the maximal blueshifted velocity of the absorption, however this evolution disappears when using the central velocity of the outflow, as shown here. Evolution in outflow velocity is also found to be more prominent at higher redshifts, with larger values of SFR and $\Sigma_{\text{SFR}}$ found to correlate with enhanced outflow velocities traced by neutral and ionised gas at $z\sim$2 \citep{sugahara17,davies19}. As such, our results are in agreement with outflow velocities found in normal galaxies at $z\sim$0, but in contrast with what is found by studies at $z\sim$2. 

Perhaps the most important quantities to look at, however, are the evolution of the (geometry-independent) mass outflow rate and mass loading factor as a function of galaxy property. Here, we are able to directly link the kpc-scale galaxy properties to the outflow by selecting spaxels with relevant galaxy properties. Using Equation \ref{eq:newdmdt1} and the mean column density per spaxel, we find log $\dot{M}_{\text{out}}$ correlates most strongly (based on the slope and range of the observed correlations) with quantities most closely related to star formation activity, however the strongest correlation is associated with log $\Sigma_{\text{SFR}}$: the mass outflow rate increases rapidly from log $\dot{M}_{\text{out}}\approx$-4 at log $\Sigma_{\text{SFR}}<$-2.25 to log $\dot{M}_{\text{out}}\sim$0 at log $\Sigma_{\text{SFR}}\sim$0.25. Milder but significant correlations are seen in the inner regions of galaxies with log sSFR, log $\Sigma_{\text{*}}$, and A$_{V}$, likely due to their own correlations with log $\Sigma_{\text{SFR}}$ and the survivability of \nad, although these all have both shallower slopes and probe smaller ranges.

Finally, we also show that the mass loading factor of the outflows again correlates most strongly with star formation activity, with values ranging -1$\lesssim$log $\eta\lesssim$1 in the central regions, and generally following the evolution of the mass outflow rate. These values and trends are similar to those found for outflowing ionised gas in star formation-driven outflows by \citep{rodriguezdelpino19} with MaNGA DR2, who also report loading factors of $\eta\lesssim$1. We are also in partial agreement with reports from \citet{davies19}, who determine strong evolution between $\eta$ and $\Sigma_{\text{SFR}}$ from ionised gas at $z\sim$2: we observe a virtually identical slope and report similar values of the loading factor, although the overlap in data is small. Thus, from this analysis we can can infer that (i) the mass outflow rate is driven primarily by the evolution of log $\Sigma_{\text{SFR}}$ and (ii) the loading factor is most closely tied to the evolution of the mass outflow rate (and hence also log $\Sigma_{\text{SFR}}$),
and (iii) that in the absence of extreme AGN, outflows are predominantly driven by the star formation activity in galaxies, traced most strongly by its surface density. Whilst such trends are clear, as mentioned in Section \ref{subsec:radial}, our ability to detect blueshifted \nad\ in absorption is highly dependent on the strength of the underlying continuum - and thus $\Sigma_{M_{*}}$ - in order to give rise to net \nad\ absorption profiles. As such, we note that some degeneracy may exist between the outflow detection limit for some galaxy properties and the associated strength of the continuum that drives the EW of the \nad\ profile. This consideration should be kept in mind when analysing the results - and in particular, the threshold detection values - from Figure \ref{fig:sfrmstarplane} and Figure \ref{fig:outflow_props}, where some underlying bias towards higher values of $\Sigma_{M_{*}}$ may be present.

\begin{figure*}
\centering
 \includegraphics[width=\textwidth]{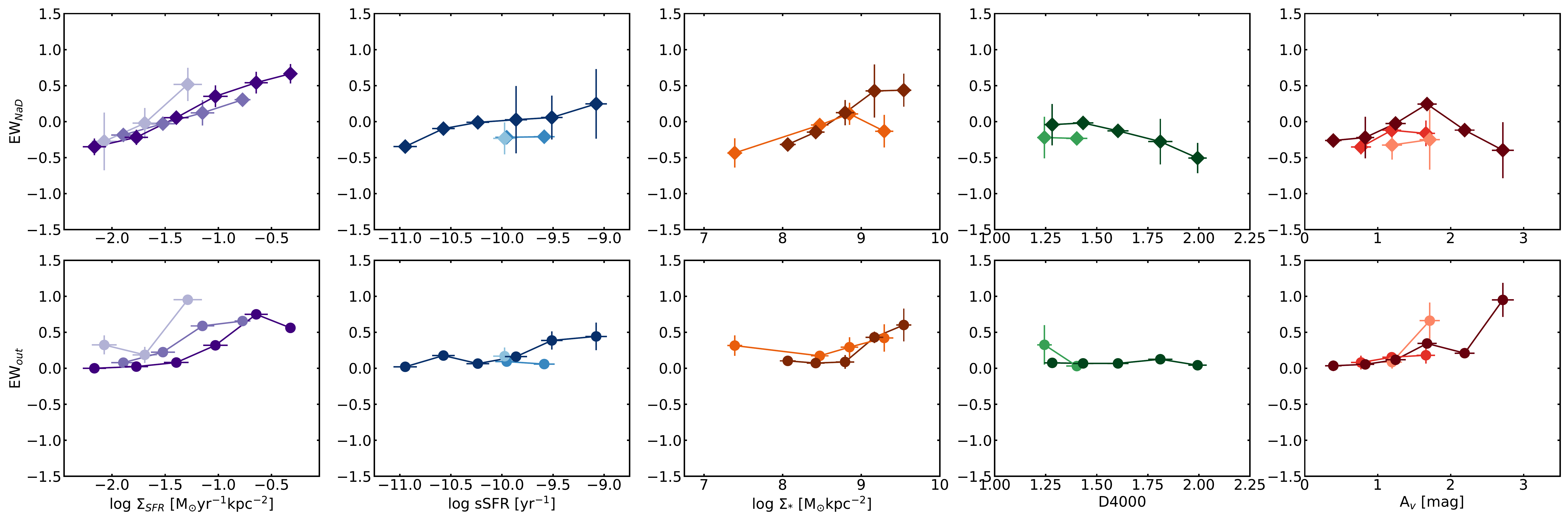}
 \caption{The evolution of the \nad\ ISM EW (top row) and blueshifted absorption EW (bottom row) as a function of $\Sigma_{\text{SFR}}$, sSFR, $\Sigma_{*}$, D(4000) and A$_{V}$. The colour scheme follows that of Figure \ref{fig:outflow_props}.}
 \label{fig:outflow_props_EW}
\end{figure*}

In addition to the above, we also look at the evolution of both the \nad\ ISM EW and the blueshifted absorption EW as a function of galaxy properties, shown in Figure \ref{fig:outflow_props_EW}. For the \nad\ ISM EW, strong positive trends become immediately clear, with the strongest trends again occurring with quantites related to star formation ($\Sigma_{\text{SFR}}$, sSFR, $\Sigma_{*}$). A tentative inverse relation is found with the D(4000) index (with older stellar populations pushing the \nad\ EW to negative values), but curiously no significant trend is found with dust, despite it being thought to regulate (in part) the survivability of \nad\ in harsh environments. Once again, the strongest of the trends is found with $\Sigma_{\text{SFR}}$ and we interpret the increase EW and \nad\ absorption as being due to the increased presence of cold gas with higher levels of $\Sigma_{\text{SFR}}$. We also note that the majority of \nad\ is found in absorption, particularly at higher values of $\Sigma_{\text{SFR}}$, sSFR, $\Sigma_{*}$ and A$_{V}$, but transforms into net emission at lower values. Interestingly, the transition of \nad\ from absorption to emission in stacks of $\Sigma_{\text{SFR}}$ occurs right around our lower detection limit at $\sim$0.01 M$_{\odot}$yr$^{-1}$kpc$^{-2}$ and consistent with what we find in the right panel of Figure \ref{fig:sfrmstarplane}. The evolution of the outflow EW as traced by blueshifted absorption is similar but less pronounced than what we observe for the entire ISM line, in the inner regions of galaxies: positive trends are found with $\Sigma_{\text{SFR}}$, sSFR and dust, with higher values corresponding to stronger outflowing gas, whilst log $\Sigma_{*}$ and $\Sigma_{*}$, D(4000) show no clear trends. Such findings are consistent with the conclusions derived above and support the notion that quantities related to star formation are the main regulators of both the \nad\ EW and outflow EW.

\section{The Impact of Outflows on HI Gas Reservoirs}
\label{sec:hicompare}
\subsection{HI Observations}
One of the most important questions in outflow studies is whether they have some impact on the cold atomic and molecular gas reservoirs of their host galaxies (i.e., can they quench the galaxy?). So far, we have inferred this from derivations of a neutral gas mass loading factor. However, an alternate approach is to examine the effect of outflows on the integrated \HI\ contents of their hosts. To do this, we make use of \HI\ 21 cm single-dish observations of galaxies in our outflow+non-detection samples with the Green Bank (GBT) and Arecibo Telescopes. 
The GBT observations form part of the HI-MaNGA \citep{masters19} program aimed at carrying out follow up \HI\ observations for MaNGA galaxies and the DR1 contains 331 galaxy observations to date. Of these, 33 are matched to our selected sample, however we also add an extra 53 galaxies which were observed after the data release (Karen Masters and Dave Stark, private communication), bringing the MaNGA GBT observations used here up to 86 galaxies. Additionally, some of our sample also overlaps with observations from the blind ALFALFA survey \citep{haynes18} conducted with the Arecibo telescope. 55 galaxies from our sample are matched to ALFALFA and as such we are able to compile \HI\ observations for a total of 141 galaxies in our combined outflow and non-detection samples, of which 34 fall under the outflow category and 107 in the non-detection category. We refer to this as our MaNGA-\HI\ sample.

\subsection{Removing the Effects of Confusion and Baseline Issues}
\label{subsec:conf}
Spectroscopic confusion in radio observations from single-dish facilities is an important concern, and in particular for stacking and outflow studies, since real signal from nearby galaxies at similar velocities can mimic the signatures of outflowing gas.

Although the rate of spectroscopic confusion is generally small (less than $\sim$15\% for the ALFALFA $\alpha$.40 data release; \citealt{jones15}), given the large beam sizes of the facilities used to obtain the \HI\ data in this study (FWHM$\approx$9$'$ and FWHM$\approx$3.5$'$ at 21cm for the GBT and Arecibo, respectively) and the potential for false-positive outflow detections, this remains a particularly important consideration.

Therefore, to ensure none of our spectra suffer from confusion, we use the MPA-JHU catalog to identify objects within 3$\times$FWHM of the beam used to observe each of our 141 \HI\ galaxies (but further than 10$''$ away from the target galaxy).
Combining this with a velocity cut of $\pm$500 \kms\ from the target source (relative to the \HI\ velocity of the target if detected, or optical redshift in the case of non-detections), we flag 78/141 of our objects as containing one or more additional optical galaxies over the search area and potentially subject to confusion. The criteria used here are extremely conservative, however we believe a rigorous approach to avoid false-positive outflow detections is crucial, and given that $\sim$80\% of the contaminating sources fall outside the FWHM of the beam (where the sensitivity drops from 50\% to effectively 0\%) combining the analysis with the beam sensitivity (see below) is likely to decrease this number significantly.

We next estimate the \HI\ gas masses of the contaminants, using gas fraction scaling relations \citep{catinella18}, and in particular the relation between log M$_{HI}$/M$_{*}$ and log sSFR. However, given that the sensitivity of the beam drops dramatically past its FWHM and we probe an area a factor of 3 larger (effectively spanning the full sensitivity of the beam) we multiply the derived \HI\ masses by the beam sensitivity at their separation in order to obtain an effective M$_{HI}$ and compare this to the target's \HI\ gas mass. If a given contaminant contributes less than 10\% to the target's \HI\ mass (in the case of a non-detection we assume upper limits as calculated in \citealt{masters19}), it is considered to have a negligible effect on the spectrum \citep{jones15}. We flag a galaxy as confused if it has one or more surrounding contaminants (within the velocity cut defined above) that contribute $\geqslant$10\% of the target's measured \HI\ gas mass. 32/141 of our MaNGA-HI sample are flagged as confused and discarded from the analysis.

A second important consideration in our stacking analysis is the possible effects resulting from badly removed baselines, which can result in artificially low or high fluxes over various regions of the spectrum. To ensure our spectra are free from such effects, we visually inspect each of the remaining 109 galaxies that are free from confusion and determine whether a baseline correction is required. 6 galaxies are found to have major baseline issues, with an additional 7 found with minor issues. For these galaxies, we fit the baseline-unsubtracted spectrum with a polynomial (generally of 3rd or 4th order) to the baseline and subtract this from the spectrum. During this procedure we consider only the regions within $\pm$1000 \kms\ of the target galaxy velocity, since fitting an accurate baseline across the entire spectrum can be challenging and our region of interest is largely confined to those velocity limits. Only 2 galaxies have major baseline issues which we are unable to correct for and an additional target contains obvious spectral artefacts, and are therefore discarded from the analysis, leaving a total of 106 galaxies free from confusion and baseline issues, which we refer to as our MaNGA-HI$_{corr}$ sample.

\subsection{Control Sample and Stacking Procedure}
\label{subsec:controlstack}
In this analysis, we wish to determine whether there is significant difference between the \HI\ gas reservoirs of galaxies with \nad\ outflows and those without. Thus, for each of the outflow galaxies in our MaNGA-HI$_{corr}$ sample, we select a control galaxy without \nad\ outflows which is matched in its position on the SFR-M$_{*}$ plane and in inclination. We allow a difference of $\pm$0.2 dex in log SFR and log M$_{*}$, as well as 20 degrees in inclination. Of our MaNGA-HI$_{corr}$ sample, we can successfully identify control galaxies for 17 of our outflow galaxies.

In order to construct mean \HI\ spectra, we stack in ``gas fraction units", that is to say we multiply each spectrum by the standard conversion factors necessary to convert to an \HI\ gas mass, described in Equation \ref{eq:mhi} and divide by the stellar mass of the galaxy. 

\begin{equation}
\label{eq:mhi}
M_{\text{\HI\ }}/\text{M}_{\odot} = 2.356\times10^{5} \bigg(\frac{D}{\text{Mpc}}\bigg)^{2} \bigg(\frac{F_{\text{HI}}}{\text{Jy}}\bigg)
\end{equation}

\begin{figure}
\center
 \includegraphics[width=0.9\columnwidth]{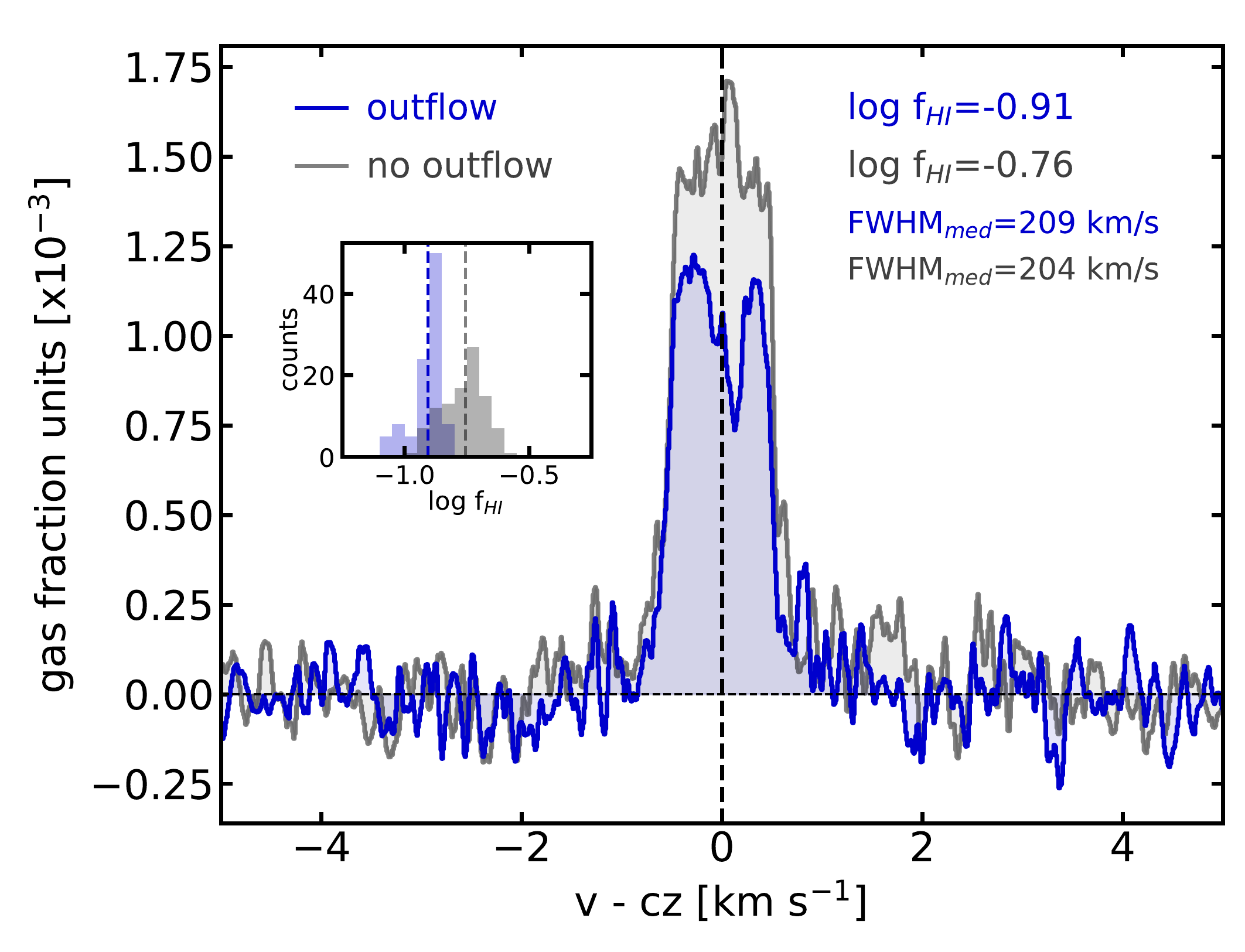}
 \caption{The mean velocity normalised, gas fraction spectra of galaxies with signatures of optical outflows (blue) and those without (gray) and their reported gas fractions and median FWHM. The stacks are created via a Monte Carlo approach of 14 outflow galaxies and a control sample of non-detection galaxies. The presence of \nad\,-selected outflows does not appear to significantly influence the \HI\ gas reservoirs of their host galaxies compared to their non-detection counterparts. The inset plot shows a histogram of the average gas fraction of a given iteration going into the final stack.}
 \label{fig:HIstack}
\end{figure}

All scaled spectra are first shifted to the rest frame prior to being normalised by the FWHM of the spectrum, then interpolated over a common velocity grid, and finally added to the stack irrespective of whether they display a clear \HI\ detection or not. The stack is then averaged to produce a mean gas fraction spectrum. We further adopt a Monte Carlo approach and repeat this process 100 times, each time with a random sample of 14 outflow galaxies ($\sim$80\% of our outflow \HI\ sample) and a different control sample, to ensure our results are not biased by a particular selection of galaxies. The final spectra are then taken as the mean over the 100 iterations and shown in Figure \ref{fig:HIstack}.

The resulting velocity-normalised spectra display significant differences between themselves, with outflow galaxies displaying reduced fluxes compared to their non-detection counterparts, suggestive of non-negligible variation in the \HI\ gas mass fractions. The mean FWHM is 209 \kms\ for the outflow galaxies and 205 \kms\ for the non-detection galaxies, with integrated \HI\ gas fractions of log $f_{\text{HI}}$=-0.906$\pm$0.001 and log $f_{\text{HI}}$=-0.757$\pm$0.001, respectively. We also present an inset plot in Figure \ref{fig:HIstack} displaying the distributions of gas fractions of the individual galaxies going into the stack, over all Monte Carlo iterations. The two distributions are distinctly offest, with means of log $f_{\text{HI}}$=-0.91 and log $f_{\text{HI}}$=-0.76 (outflow detections and non-detections, respectively) consistent with the integrated spectra, and spreads characterised by standard deviations of log $\sigma_{f_{\text{HI}}}\approx$-1.79 (outflow detections) and log $\sigma_{f_{\text{HI}}}\approx$-1.47 (non-detections). The two distributions reinforce the apparent lack of similarity between the two stacked spectra and the combination of the two is suggestive of a non-negligible difference between the \HI\ gas fractions of the outflow-selected galaxies and those without outflows. 

Although causality is challenging to establish, we have been careful to match the outflow-detection and non-detection galaxies in several key properties that could influence the observed \HI\ gas fraction, to ensure the primary difference is the detection of an optical outflow. Thus, the different shapes of the profiles (a double horn profile for outflow galaxies and a Gaussian-like profile for the non-detection galaxies) could hint at blowing out of \HI\ gas at the centres of the galaxies by the outflows (i.e., at normalised velocities of $\sim$0 \kms), where the outflows are strongest (see Section \ref{subsec:radial}). Furthermore, the higher \HI\ gas fractions at virtually all velocities in the non-detection stack are suggestive of some gas replenishment and could provide indirect evidence for the interplay between outflows and inflows - i.e., a galactic fountain scenario. Whilst, such a scenario is a distinct possibility, we also note that such features could also reflect some transition of neutral \HI\ gas into molecular or ionised gas, in which case more molecular gas may be available for star formation in the galaxy centres and could lead to the formation of the winds. However, distinguishing between and confirming such scenarios is beyond the capabilities of the current data sets.


\section{Discussion}
\label{sec:discuss}
\subsection{Strong Correlations With $\Sigma_{\text{SFR}}$}
Throughout each of the previous Sections, we have seen that $\Sigma_{\text{SFR}}$ plays a key role in regulating the output of outflows in normal galaxies at $z\sim$0. Whilst other quantities (e.g., sSFR, $\Sigma_{*}$ and A$_{V}$) also correlate with outflow quantities, these appear secondary to the dominant correlations found with $\Sigma_{\text{SFR}}$ and likely exist due to their own correlations with $\Sigma_{\text{SFR}}$. Similar correlations have also been observed in other studies, both at low and high redshifts \citep{chen10,newman12a,rb18,davies19}. For galaxies at $z\sim$2, \citet{newman12a} suggested the possibility of a critical ``blow out" $\Sigma_{\text{SFR}}$, where star formation feedback is able to generate enough pressure to perpendicularly break out of the dense gas of the disk in the form of a momentum-driven outflow. By assuming a baryon dominated galaxy disk that sits in pressure equilibrium, one can test this via simple assumptions that equate the weight of the disk gas to the pressure exerted outward by star formation feedback, and determine a minimum $\Sigma_{\text{SFR}}$ threshold above which pressure from the feedback exceeds the weight of the gas.

Following Equations 1 and 7 from \citet{ostriker11}, the weight of the disk can be expressed as

\begin{equation}
w = \frac{\pi G \Sigma_{\text{gas}}^{2}}{2},
\end{equation}

where $\Sigma_{\text{gas}}$ is the cold gas density and $G$ is the gravitational constant. The vertical momentum flux injected into the ISM by stellar feedback can be described as

\begin{equation}
P = \frac{p_{*}}{4m_{*}}\Sigma_{\text{SFR}},
\end{equation}

where ($p_{*}$/$m_{*}$) is the mean radial momentum injected into the ISM per unit mass of stars formed. As described by Equation 2 of \citet{newman12a}, one can rearrange these two equations and insert a dependence on gas fraction, $f_{g}$ and stellar mass density to derive a minimum $\Sigma_{\text{SFR}}$ threshold. This is particularly useful given that such quantities are well constrained for local, normal galaxies and by the data presented here. As such, our final equation is

\begin{equation}
\Sigma_{\text{SFR,thresh}} = \frac{\pi G f_{\text{g}}}{2(p_{*}/m_{*})} \Sigma_{*}^{2}.
\end{equation}

Assuming $f_{g}\sim$0.5, ($p_{*}$/$m_{*}$)$\sim$1000 \kms and $\Sigma_{\text{gas}}\sim$500-1000 M$_{\odot}$pc$^{-2}$ for their sample of normal SFGs at $z\sim$2, \citet{newman12a} found a critical threshold of $\sim$1 M$_{\odot}$yr$^{-1}$kpc$^{-2}$, in good agreement with their observations of ionised outflows. Assuming typical values ($f_{g}\sim$0.07, $\Sigma_{\text{gas}}\sim$500) for $z\sim$0 normal SFGs, this threshold is lowered down to $\sim$0.1 M$_{\odot}$yr$^{-1}$kpc$^{-2}$, in agreement with the canonical value typically assumed as the minimum threshold to launch outflows \citep{heckman02, veilleux05, ostriker11}.

However, this threshold is typically observed in more turbulent, starburst systems and not in normal SFGs along the MS. Here, we expand on this by using values directly inferred from our MaNGA data set. Assuming a cold (molecular) gas fraction for high mass galaxies of $f_{g}\sim$0.04 \citep{saintonge17} and a mean, galaxy-wide mass surface density of log $\Sigma_{*}$/M$_{\odot}$kpc$^{-2}\sim$8.24 over all of our sample of 376 outflow+non-detection galaxies, we derive a critical threshold of $\Sigma_{\text{SFR}}\sim$0.01 M$_{\odot}$yr$^{-1}$kpc$^{-2}$, in agreement with the results presented here and in \citet{rb18}. Additionally, we also note that the mean galaxy-wide $\Sigma_{\text{SFR}}$ for our outflow selected galaxies is $\sim$0.02 M$_{\odot}$yr$^{-1}$kpc$^{-2}$ whilst that of the non-detection galaxies is $\sim$0.01 M$_{\odot}$yr$^{-1}$kpc$^{-2}$, sitting above and on our derived threshold, respectively.

For reference, we compare this value to the average SFR surface density of the Milky Way, which does not harbour a strong outflow \citep{fox19}. Assuming a SFR of 1.65 M$_{\odot}$\,yr$^{-1}$ (\citealt{bh2016}, and references therein) and a disk radius of 10 kpc (\citealt{olsen15}, and references therein), the area of the disk is 314 kpc$^{2}$ and thus the SFR surface density is calculated to be $\Sigma_{\text{SFR}}\sim$0.005 M$_{\odot}$yr$_{-1}$kpc$^{-2}$, below our derived threshold. Such values can provide a useful comparative framework, however similar analyses over the central regions of the Milky Way (e.g., the Central Molecular Zone or the Galactic Centre) could provide a more informative comparison given they likely provide the regions with the strongest outflowing gas, whilst the star formation history of these regions should also be taken into account.

\subsection{Star Formation Histories of Outflow Hosts}

Given that star formation quantities appear to be the primary drivers of our selected outflows, it is instructive to look at the star formation histories associated with the host galaxies which give rise to the outflows. As such, we use the D$_{n}$(4000) and H$\delta_{\text{A}}$ indices provided by the MaNGA DAP, whose combination can reveal and distinguish between continuous and bursty star formation histories \citep{kauffmann03}. The D$_{n}$(4000) index measures the strength of the 4000\,\AA\ absorption break, which is a discontinuity in the optical spectrum of galaxies due to a variety of absorption features primarily arising from ionised metals in the atmospheres of stars. The index is linked to the age of the galaxy's stellar populations, given that young, hot stars ionise the surrounding metals, leading to a decreased opacity, whilst this is not the case for old, metal rich stars which display considerably more metal absorption. As such, a large 4000\,\AA\ break is indicative of older stellar populations, whilst a smaller one is indicative of younger stellar populations and a generally increasing break is found with time, as the stellar populations evolve. The index itself is measured as the ratio of the average continuum flux in regions blue and redward of the break, with several sources in the literature adopting differently-sized regions (broad or narrow). The advantage of adopting a more narrow-sized region either side of the break is that the ratio is less sensitive to the effects of dust. As such, we use the narrow-band definition adopted by MaNGA (3850-3950 and 4000-4100\,\AA\ for the blue and red continuum bandpasses, respectively; \citealt{westfall19}). 

H$\delta$ absorption at 4101\,\AA, on the other hand, arises from Balmer absorption in the atmospheres of stars and is an indication of the timing of a recent burst of star formation. Beginning with weak intrinsic absorption by hot O and B stars (characterised by short lifetimes of a few 10 Myrs) at the time of the burst, the absorption increases monotonically over time until $\sim$300 Myr where it reaches its peak. The rapid increase and peak of the evolution is due to the deaths of the OB stars and domination of the optical light by late-B to early-F stars (characterised by longer lifetimes of $\lesssim$1 Gyr). After the peak absorption at $\sim$300 Myr, the H$\delta$ absorption rapidly decreases as the A and F stars die off. As such, H$\delta$ absorption is a measure of recent bursts of star formation that ended only $\sim$0.1-1 Gyr \citep{kauffmann03}, and its index (H$\delta_{A}$) is measured using a central bandpass marked by two pseudo-continuum bandpasses either side of the line (4083-4122, 4041-4079, 4128-4161\,\AA\ for the main, blue and red bandpasses, respectively; \citealt{westfall19}).

The two indices are largely independent of reddenning effects, however metallicity can play a role in regulating them at later times. However, as shown by \citet{kauffmann03}, metallicity only becomes an important consideration for both tracers at ages of $>$10$^{9}$ years after a burst (corresponding to rough values of D$_{n}$4000$>$1.5 and H$\delta_{\text{A}}<$3-4).

\begin{figure}
\center
 \includegraphics[width=0.99\columnwidth]{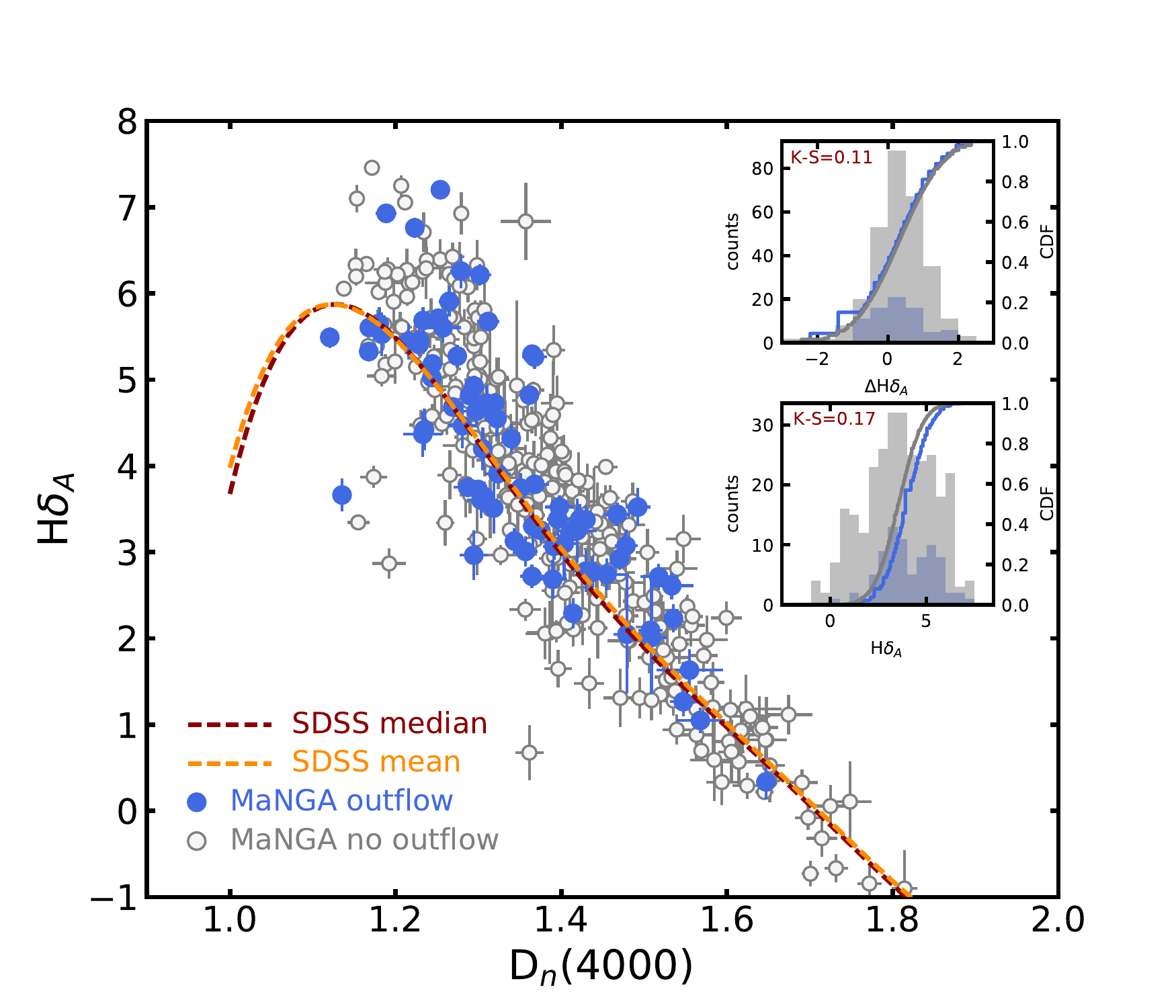}
 \caption{The star formation histories as traced by the 3$''$-diameter stacked H$\delta_{\text{A}}$ and D$_{n}$(4000) indices of our MaNGA DR15 sample of galaxies shown to host \nad\ outflows (blue points) and those that do not (gray points). The error bars represent the standard deviation of values going into the stack for each galaxy. These points are compared to the mean (dashed orange line) and median (dashed red line) of SDSS values from the MPA-JHU catalogs. The inset plots show the distributions of $\Delta$H$\delta_{A}$ (H$\delta_{A,\text{MaNGA}}$-H$\delta_{A,\text{SDSS}}$; top) and H$\delta_{A}$ (bottom) for outflows (blue lines) and non-detections (gray lines), as well as their cumulative distribution functions and associated two-sample Kolmogorov-Smirnov statistic.}
 \label{fig:SFhist}
\end{figure}

To conduct our analysis, we use the average of the two tracers within a galactocentric radius of 1.5$''$ for each of our outflow and non-detection galaxies and compare these to results in the MPA-JHU catalog as measured by the SDSS DR7 3$''$-diameter fiber. The SDSS relations are taken by binning star-forming galaxies selected by the same MaNGA selection criteria described in Section \ref{sec:stacking} in steps of $\Delta$D$_{n}$(4000)=0.2 and taking the mean and median values, before fitting a 5th order polynomial. We present these fits and data points in Figure \ref{fig:SFhist}.

We find our data points span a wide range over both indices and generally sit slightly above the mean (and median) SDSS relation, regardless of whether they are selected to have \nad\ outflows or not. For the outflow galaxies, $\sim$63\% of the data points sit above the mean SDSS relation and $\sim$37\% below, suggesting the majority of the data points have experience higher bursts of star formation in their recent past. However, comparing this to the non-detection data points, which reveal a slightly higher percentage of $\sim$70\% sitting above the SDSS relation and $\sim$30\% sitting below, we find very similar values. This is in contrast to IFU observations of ionised tracers in edge-on galaxies found by \citet{ho-i16}, who observe 80\% of their outflow galaxies and 50\% of their non-detection galaxies to sit above the median SDSS relation, albeit with sample sizes a factor of $\sim$5 and $\sim$12 smaller than those used here, respectively. We expand on this by comparing the distribution and mean values of our data points with the corresponding SDSS values ($\Delta$H$\delta_{A}$=H$\delta_{A,\text{MaNGA}}$-H$\delta_{A,\text{SDSS}}$), inferred from our fit relation. We present this in the top inset plot in Figure \ref{fig:SFhist}. The two distributions are very similar, with outflows characterised by a mean difference of 0.27 and standard deviation of 0.78, and non-detections with a mean of 0.32 and identical standard deviation. This is reinforced by near identical cumulative distribution functions (CDFs) and a low Kolmogorov-Smirnov statistic of 0.11, suggesting little difference between the two distributions and supporting the notion that both the outflow and non-detection samples have both experienced more recent star formation bursts compared to their SDSS counterparts. Such similarity extends to the two MaNGA samples, which we compare via their H$\delta_{A}$ distributions in the bottom inset plot in Figure \ref{fig:SFhist}. The two distributions are once again quite similar, with means of 4.02 and 3.56 and standard deviations of 1.42 and 1.82 for the outflow and non-detection samples, respectively. The two CDFs and a K-S statistic of 0.17 reinforce this similarity, albeit with a slight offset for outflow galaxies towards higher H$\delta_{A}$ values. We do note, however, that the distribution of the non-detection galaxies does fall down to H$\delta_{A}<$3-4, where the tracer becomes uncertain due to metallicity effects, and the differences in mean values is attributed to this extension.

The similarities between the two MaNGA samples suggest both the outflow-detected and non-detected samples may be undergoing similar episodes of short bursts of more intense star formation that may be capable of launching outflows. If this is the case, both samples would display similar values of H$\delta_{A}$ and D$_{n}$(4000), although one may also expect more of the non-detection galaxies to display signatures of optical outflows. Thus, the lack of (detected) outflow signatures in the non-detection sample is perhaps surprising. Following on from this, whilst our results are in agreement with the ionised gas results of \citet{ho-i16} in that the majority of our outflow data points lie above the mean SDSS relation, the considerable similarity between our larger distributions of outflow and non-detection samples is in contrast to theirs and suggests that (i) our two MaNGA samples may be undergoing similar episodes of star formation and (ii) that the timing of the last burst of star formation may not be a crucial parameter in driving (weak) \nad\ outflows.



\section{Summary and Conclusions}
\label{sec:concl}
We use the SDSS-IV/MaNGA DR15 data set and spectral stacking techniques to constrain the main kpc-scale properties that give rise to and regulate neutral outflows in star-forming galaxies. We use a sample of 405 $z\sim$0 face-on galaxies along the galaxy MS to determine the detection fraction, galactocentric radial profile and kpc-scale properties of galaxies with signatures of blueshifted \nad\ absorption. Our main findings can be summarised as follows:

\begin{itemize}
\item Out of 390 useable galaxies in our sample, the stacking of the central 0.25\,R$_{e}$ regions of the galaxies reveals 78 objects with signatures of outflowing gas, 14 objects with signatures of inflowing gas, and 298 galaxies with no significant detection of blue/redshifted gas.
Galaxies with signatures of outflows and inflows show considerably higher values of $\Sigma_{*}$ and $\Sigma_{\text{SFR}}$ compared to their non-detection counterparts.
\\
\item Stacking as a function of deprojected galactocentric radius, we find detections of outflowing gas out to 1\,R$_{e}$ and compare these to detections as observed by an 3$''$-diameter fiber. Derivations of a mass outflow rate show a rapidly declining trend with galactocentric radius, with a range -2.1$\lesssim$log $\dot{M}_{\text{out}} \lesssim$-0.9 and best fit slope of -1.2. A near identical trend is found with the mass loading factor, which also decreases as a function of galaxy radius, demonstrating that outflows are at their strongest in the central regions ($<$0.5\,R$_{e}$) of galaxies. The range of reported mass loading factors is -1.8$\lesssim$log $\eta \lesssim$-0.4, consistent with previously reported values in relatively normal galaxies at $z\sim$0, and a slope of -1.4. We provide prescriptions for both of these trends for use in simulations.
\\
\item Signatures of outflowing gas are found along and above the resolved star-forming MS, in parameter space above log $\Sigma_{*} \gtrsim$7.5 M$_{\odot}$kpc$^{-2}$ and log $\Sigma_{\text{SFR}} \gtrsim$-2 M$_{\odot}$yr$^{-1}$kpc$^{-2}$, similar to trends found along the galaxy MS. A comparison of the outflow EW, mass outflow rate and mass loading factor shows an increase in each of these values as one moves up the resolved MS, suggestive of stronger outflows in regions of increased star formation activity.
\\
\item By stacking $>$57,500 individual spaxels associated with a variety of galaxy properties ($\Sigma_{\text{SFR}}$, $\Sigma_{*}$, specific SFR, D(4000) and A$_{V}$), we find significant positive correlations between outflow properties (namely, gas column density, mass outflow rate and mass loading factor) and all of the aforementioned galaxy properties, except D(4000). The strongest correlations are found with log $\Sigma_{\text{SFR}}$ and we extend the lower limit of detections down to log $\Sigma_{\text{SFR}}\approx$-2 M$_{\odot}$yr$^{-1}$kpc$^{-2}$, about an order of magnitude lower than canonical values. We suggest that this is due to a minimum $\Sigma_{\text{SFR}}$ threshold of $\Sigma_{\text{SFR}}\approx$0.01 M$_{\odot}$yr$^{-1}$kpc$^{-2}$ necessary for star formation feedback to break out of the dense gas in the disk.
\\
\item Using \HI\ follow up observations of a sample of 106 MaNGA galaxies, we compare the cold \HI\ gas reservoirs of galaxies selected to have \nad\ outflows to control samples of galaxies without \nad\ outflows, matched by their position in the SFR-M$_{*}$ plane and inclination. Stacking the \HI\ spectra, we find considerable differences between gas reservoirs of outflow and non-detection galaxies, suggesting the presence of optically-selected outflowing gas could potentially have some (minor) effect on the cold gas reservoirs, although combining this with our \nad\ mass loading factors strongly suggests that it is not enough for normal galaxies to be quenched by ejective feedback.
\\
\item Finally, we compare the star formation histories of our outflow and non-detection galaxies, using the central (3$''$) stacked H$\delta_{A}$ and D$_{n}$(4000) indices. In comparing to average SDSS DR7 values, we find our sample of MaNGA galaxies have slightly elevated values of H$\delta_{A}$ compared to the SDSS galaxies for a given D$_{n}$(4000), however, crucially we find virtually no difference between the values of our outflow sample and non-detection sample, suggesting that the timing since the last burst of star formation is not a crucial parameter in driving \nad\ outflows.
\end{itemize}

The arrival of large IFU surveys has greatly aided studies of local galaxy evolution and, in particular, studies of outflows. The kinematics and key properties of outflows and their hosts are being studied in unprecedented detail and the small, kpc-scale processes that give rise to the large scale processes are being constrained. Here, we have used samples of star-forming galaxies and $>$275,000 galaxy resolved spectra to determine the prevalence of outflows on kpc-scales, their power and quenching potential across a variety of galaxy regions and properties and determined whether any significant impact on the \HI\ gas reservoirs in seen. However, still missing are crucial constraints on the multiphase nature of outflows in normal, MS galaxies at $z\sim$0 in order to infer their true quenching potential. Verifying this will require rest-frame optical, (sub)millimetre and radio spectra over statistically meaningful samples of galaxies with which to probe and compare the different outflowing gas phases, with the aim of deriving total mass loading factors over large regions of parameter space. Large surveys such as the SDSS, xCOLD GASS and ALFALFA provide unique opportunities to do this, whilst the arrival of optical and radio surveys such as MaNGA, SAMI and PHANGS will allow similar analyses to be achieved on resolved scales.

\section*{Acknowledgements}
This research was supported by grants from the Royal Society. 
The authors would like to thank the referee, Sylvain Veilleux, for a helpful and thoughtful report which significantly improved this paper. GRB would also like to thank Barbara Catinella for useful discussions on criteria for establishing \HI\ confusion, as well as Thomas Greve and Marc Sarzi for useful comments that improved this paper.
This paper also makes use of data from the Green Bank Telescope. The Green Bank Observatory is a facility of the National Science Foundation operated under cooperative agreement by Associated Universities, Inc.

Funding for the Sloan Digital Sky Survey IV has been provided by the Alfred P. Sloan Foundation, the U.S. Department of Energy Office of Science, and the Participating Institutions. SDSS acknowledges support and resources from the Center for High-Performance Computing at the University of Utah. The SDSS web site is www.sdss.org. \\
SDSS is managed by the Astrophysical Research Consortium for the Participating Institutions of the SDSS Collaboration including the Brazilian Participation Group, the Carnegie Institution for Science, Carnegie Mellon University, the Chilean Participation Group, the French Participation Group, Harvard-Smithsonian Center for Astrophysics, Instituto de Astrof\'isica de Canarias, The Johns Hopkins University, Kavli Institute for the Physics and Mathematics of the Universe (IPMU) / University of Tokyo, the Korean Participation Group, Lawrence Berkeley National Laboratory, Leibniz Institut für Astrophysik Potsdam (AIP), Max-Planck-Institut für Astronomie (MPIA Heidelberg), Max-Planck-Institut für Astrophysik (MPA Garching), Max-Planck-Institut für Extraterrestrische Physik (MPE), National Astronomical Observatories of China, New Mexico State University, New York University, University of Notre Dame, Observatório Nacional / MCTI, The Ohio State University, Pennsylvania State University, Shanghai Astronomical Observatory, United Kingdom Participation Group, Universidad Nacional Autónoma de México, University of Arizona, University of Colorado Boulder, University of Oxford, University of Portsmouth, University of Utah, University of Virginia, University of Washington, University of Wisconsin, Vanderbilt University, and Yale University.


\appendix
\section{Global Properties of Flow Galaxies}
\label{sec:appendixa}
We present here the main global properties of MaNGA DR15 galaxies identified to host either an outflow or inflow in the central regions of the galaxy, whose properties were used in the sample selection. Quantities with superscript $a$ mark quantities derived from the NSA catalog, whilst quantities with superscript $b$ are derived from the Pipe3D catalog. SFRs and stellar masses are integrated quantities, whilst the line ratios refer to measurements in the central 2.5$''$ of the galaxy.

\onecolumn
\begin{longtable}{@{}ccccccccccc@{}}
\caption{The global properties of galaxies identified to host outflows in their central regions.}
\label{tab:samplegals}
\\ \hline
Plate ID & IFU & RA\tablenotemark{a} & DEC\tablenotemark{a} & $z$\tablenotemark{a} & log M$_{*}$\tablenotemark{b} & log SFR\tablenotemark{b} & log [\OIII]/H$\alpha$\tablenotemark{b} & log [\NII]/H$\beta$\tablenotemark{b} & $i$\tablenotemark{a} & Type \\
         &     & [deg]    & [deg]     &     & [M$_{\odot}$] & [M$_{\odot}$\,yr$^{-1}$] &  &  & [deg] & \\
\hline
7815 & 3702 & 317.90320 & 11.49694 & 0.02938 & 10.47 & 0.43 & -0.73 & -0.36 & 27.49 & outflow \\
8077 & 3702 & 41.84637 & 0.05876 & 0.02478 & 10.12 & -0.10 & -0.59 & -0.37 & 23.92 & outflow \\
8085 & 3704 & 51.15552 & -0.44402 & 0.03684 & 10.72 & 0.66 & -0.74 & -0.44 & 44.23 & outflow \\
8134 & 1901 & 113.40018 & 45.94338 & 0.07665 & 10.98 & 1.29 & -0.63 & -0.49 & 27.01 & outflow \\
8146 & 1901 & 117.05387 & 28.22509 & 0.02706 & 10.38 & 0.42 & -0.69 & -0.43 & 22.85 & outflow \\
8147 & 6103 & 119.03950 & 26.87578 & 0.06161 & 11.06 & 1.00 & -0.72 & -0.48 & 36.35 & outflow \\
8149 & 3704 & 120.20722 & 27.50036 & 0.01734 & 10.30 & 0.62 & -0.55 & -0.43 & 32.84 & outflow \\
8155 & 3703 & 54.09596 & -0.88266 & 0.02358 & 10.18 & 0.27 & -0.11 & -0.52 & 25.01 & outflow \\
8253 & 3703 & 157.34358 & 43.17059 & 0.02726 & 10.32 & 0.39 & -0.72 & -0.47 & 45.99 & outflow \\
8254 & 1902 & 163.21911 & 43.42840 & 0.02424 & 10.09 & 0.02 & -0.81 & -0.37 & 27.02 & outflow \\
8262 & 9102 & 184.55357 & 44.17324 & 0.02453 & 10.66 & 0.72 & -0.80 & -0.34 & 39.23 & outflow \\
8309 & 6104 & 211.78062 & 52.96374 & 0.04333 & 10.21 & 0.33 & -0.50 & -0.53 & 38.35 & outflow \\
8311 & 3703 & 205.01217 & 23.14297 & 0.03168 & 10.84 & 0.60 & -0.40 & -0.34 & 34.63 & outflow \\
8312 & 12702 & 245.27087 & 39.91739 & 0.03203 & 10.70 & 0.52 & -0.51 & -0.30 & 43.71 & outflow \\
8312 & 12703 & 247.20906 & 39.83509 & 0.03585 & 10.83 & 0.73 & -0.46 & -0.36 & 16.04 & outflow \\
8313 & 12701 & 239.49180 & 41.79259 & 0.03547 & 10.78 & 0.72 & -0.52 & -0.45 & 47.27 & outflow \\
8313 & 12702 & 240.67742 & 41.19726 & 0.03334 & 10.73 & 1.06 & -0.69 & -0.49 & 48.87 & outflow \\
8317 & 6102 & 194.92503 & 43.75318 & 0.05781 & 11.13 & 1.16 & -0.36 & -0.39 & 35.55 & outflow \\
8329 & 6104 & 213.11048 & 45.69041 & 0.02700 & 10.54 & 0.73 & -0.43 & -0.32 & 31.33 & outflow \\
8332 & 3702 & 207.87281 & 43.80643 & 0.03331 & 10.56 & 0.67 & -0.28 & -0.39 & 39.89 & outflow \\
8333 & 3701 & 214.34452 & 42.28797 & 0.06861 & 10.84 & 0.72 & -0.40 & -0.34 & 27.82 & outflow \\
8446 & 6104 & 206.54141 & 36.50439 & 0.05524 & 10.54 & 0.84 & -0.52 & -0.46 & 37.46 & outflow \\
8454 & 12702 & 153.04902 & 45.14216 & 0.07643 & 11.13 & 1.40 & -0.33 & -0.36 & 44.77 & outflow \\
8454 & 6104 & 154.82544 & 45.81359 & 0.06348 & 11.07 & 1.12 & -0.47 & -0.33 & 34.15 & outflow \\
8455 & 3701 & 157.17946 & 39.83888 & 0.02929 & 10.34 & 0.60 & -0.58 & -0.38 & 41.65 & outflow \\
8465 & 12705 & 198.23632 & 47.45663 & 0.02814 & 10.81 & 0.11 & -0.44 & -0.34 & 47.01 & outflow \\
8465 & 9102 & 198.18916 & 46.93499 & 0.02768 & 10.40 & 0.12 & -0.83 & -0.47 & 33.48 & outflow \\
8486 & 6101 & 238.03955 & 46.31979 & 0.05887 & 10.88 & 0.60 & -0.44 & -0.33 & 43.53 & outflow \\
8550 & 12703 & 247.67443 & 40.52939 & 0.02981 & 10.49 & 0.05 & -0.65 & -0.33 & 31.70 & outflow \\
8550 & 9102 & 247.20906 & 39.83509 & 0.03585 & 10.81 & 0.73 & -0.46 & -0.35 & 16.04 & outflow \\
8551 & 12705 & 233.94092 & 44.83480 & 0.02964 & 10.59 & 0.79 & -0.56 & -0.40 & 26.90 & outflow \\
8554 & 3704 & 184.62322 & 35.62226 & 0.03490 & 10.44 & 0.35 & -0.60 & -0.45 & 29.22 & outflow \\
8588 & 6101 & 248.45676 & 39.26321 & 0.03176 & 10.88 & 0.67 & -0.41 & -0.34 & 16.10 & outflow \\
8600 & 1901 & 242.58522 & 43.00964 & 0.02521 & 10.19 & 0.48 & -0.16 & -0.58 & 46.99 & outflow \\
8611 & 3702 & 261.46394 & 60.19412 & 0.02905 & 10.51 & 0.13 & -0.53 & -0.35 & 41.79 & outflow \\
8616 & 3703 & 322.51023 & 0.46420 & 0.13506 & 11.15 & 1.21 & -0.56 & -0.38 & 28.05 & outflow \\
8618 & 3704 & 318.86229 & 9.75782 & 0.07023 & 10.94 & 0.82 & -0.59 & -0.41 & 10.25 & outflow \\
8624 & 9102 & 263.89258 & 59.88992 & 0.02836 & 10.55 & 0.27 & -0.54 & -0.36 & 48.19 & outflow \\
8626 & 12703 & 265.10878 & 56.70125 & 0.06928 & 10.92 & 0.66 & -0.34 & -0.33 & 41.12 & outflow \\
8626 & 3703 & 264.66254 & 56.82424 & 0.02942 & 10.33 & 0.51 & -0.45 & -0.44 & 38.56 & outflow \\
8715 & 9101 & 119.91012 & 51.79236 & 0.05338 & 10.72 & 0.63 & -0.63 & -0.35 & 48.18 & outflow \\
8725 & 3703 & 125.82096 & 45.89686 & 0.05373 & 10.64 & 0.59 & -0.59 & -0.34 & 44.09 & outflow \\
8725 & 6102 & 125.45904 & 45.51964 & 0.05417 & 10.66 & 0.87 & -0.66 & -0.33 & 22.84 & outflow \\
8940 & 1902 & 120.98713 & 25.48072 & 0.07278 & 11.10 & 1.29 & -0.60 & -0.38 & 29.88 & outflow \\
8945 & 3702 & 173.36190 & 47.28673 & 0.04555 & 10.05 & 0.65 & 0.07 & -0.67 & 29.13 & outflow \\
8977 & 9102 & 119.20025 & 33.25448 & 0.06756 & 10.86 & 1.16 & -0.47 & -0.46 & 35.26 & outflow \\
8992 & 12704 & 172.63126 & 51.45594 & 0.02559 & 10.55 & 0.48 & -0.77 & -0.52 & 40.50 & outflow \\
9026 & 3703 & 249.34435 & 44.34712 & 0.03069 & 10.48 & 0.66 & -0.87 & -0.51 & 19.68 & outflow \\
9028 & 9101 & 243.02559 & 28.49887 & 0.05318 & 10.83 & 0.84 & -0.84 & -0.45 & 27.34 & outflow \\
9029 & 12702 & 247.25172 & 41.28426 & 0.03189 & 10.80 & 0.57 & -0.66 & -0.49 & 33.56 & outflow \\
9029 & 1902 & 246.48724 & 41.52207 & 0.04272 & 10.03 & 0.01 & -0.74 & -0.32 & 22.29 & outflow \\
9033 & 1902 & 223.21854 & 45.23480 & 0.13265 & 11.16 & 1.27 & -0.55 & -0.41 & 37.88 & outflow \\
9033 & 3703 & 222.93490 & 47.86576 & 0.04988 & 10.56 & 0.53 & -0.45 & -0.35 & 27.50 & outflow \\
9035 & 6102 & 236.65444 & 43.68365 & 0.10117 & 11.03 & 1.18 & -0.55 & -0.40 & 47.29 & outflow \\
9041 & 9102 & 235.94192 & 28.41521 & 0.03275 & 10.66 & 0.67 & -0.55 & -0.46 & 33.55 & outflow \\
9042 & 12703 & 235.15268 & 28.51244 & 0.03275 & 10.71 & 0.23 & -0.62 & -0.37 & 46.84 & outflow \\
9044 & 6101 & 230.68698 & 29.76960 & 0.02292 & 10.24 & 0.39 & -0.78 & -0.47 & 33.96 & outflow \\
9049 & 9102 & 248.38654 & 25.96446 & 0.04532 & 10.33 & 0.35 & -0.79 & -0.40 & 18.95 & outflow \\
9088 & 12704 & 242.74641 & 26.34917 & 0.06459 & 10.95 & 0.59 & -0.53 & -0.39 & 33.25 & outflow \\
9095 & 1901 & 242.81005 & 24.22502 & 0.03253 & 10.62 & 0.76 & -0.72 & -0.32 & 40.78 & outflow \\
9182 & 1902 & 120.64756 & 38.74119 & 0.11639 & 11.13 & 1.21 & -0.86 & -0.38 & 33.58 & outflow \\
9195 & 1902 & 28.60509 & 12.66317 & 0.04466 & 10.31 & 0.59 & -0.51 & -0.40 & 22.21 & outflow \\
9195 & 6104 & 28.96659 & 14.94028 & 0.04382 & 11.00 & 1.12 & -0.59 & -0.48 & 44.11 & outflow \\
9485 & 1901 & 120.77803 & 37.02356 & 0.07122 & 10.96 & 1.40 & -0.62 & -0.41 & 16.15 & outflow \\
9487 & 12703 & 125.08855 & 45.53436 & 0.05413 & 10.68 & 0.79 & -0.59 & -0.41 & 45.79 & outflow \\
9490 & 3704 & 122.51965 & 19.74559 & 0.04485 & 10.32 & 0.34 & -0.27 & -0.46 & 33.68 & outflow \\
9491 & 12704 & 119.53574 & 19.55015 & 0.06159 & 10.81 & 0.88 & -0.82 & -0.50 & 41.37 & outflow \\
9491 & 6101 & 119.17438 & 17.99117 & 0.04124 & 10.90 & 1.30 & -0.28 & -0.34 & 46.30 & outflow \\
9493 & 9102 & 130.93325 & 22.96240 & 0.06155 & 10.82 & 0.81 & -0.60 & -0.40 & 32.95 & outflow \\
9494 & 3704 & 128.01990 & 21.62863 & 0.05376 & 10.75 & 0.95 & -0.53 & -0.31 & 44.86 & outflow \\
9506 & 3701 & 133.56990 & 27.26653 & 0.06378 & 11.06 & 1.40 & -0.70 & -0.46 & 24.30 & outflow \\
9508 & 6101 & 127.13102 & 26.34868 & 0.05303 & 10.59 & 1.30 & -0.39 & -0.37 & 30.04 & outflow \\
9509 & 1901 & 122.38423 & 26.28342 & 0.02542 & 10.14 & 0.26 & -0.54 & -0.43 & 37.66 & outflow \\
9869 & 9102 & 247.25661 & 40.53487 & 0.02785 & 10.51 & 0.04 & -0.52 & -0.31 & 49.73 & outflow \\
9871 & 12705 & 228.37217 & 42.33025 & 0.02795 & 10.63 & 0.37 & -0.68 & -0.51 & 38.57 & outflow \\
9881 & 3702 & 203.81165 & 25.04475 & 0.02609 & 10.40 & 0.31 & -0.70 & -0.40 & 31.59 & outflow \\
9883 & 9102 & 256.64139 & 33.69298 & 0.03015 & 10.45 & 0.48 & -0.60 & -0.33 & 46.58 & outflow \\
9888 & 12705 & 236.90138 & 26.06377 & 0.03158 & 10.94 & 1.00 & -0.55 & -0.38 & 43.63 & outflow \\
7977 & 9102 & 332.83066 & 12.18472 & 0.06309 & 10.93 & 0.83 & -0.42 & -0.35 & 49.52 & inflow \\
8081 & 9101 & 47.77218 & -0.54654 & 0.02820 & 10.60 & 0.32 & -0.32 & -0.36 & 49.93 & inflow \\
8084 & 3702 & 50.63664 & -0.00121 & 0.02174 & 10.23 & 0.43 & -0.14 & -0.39 & 41.58 & inflow \\
8322 & 1901 & 198.78425 & 30.40377 & 0.02315 & 10.38 & 0.51 & -0.58 & -0.33 & 34.58 & inflow \\
8602 & 3701 & 247.36154 & 38.41940 & 0.03053 & 10.18 & -0.11 & -0.59 & -0.38 & 44.22 & inflow \\
8717 & 3704 & 117.51833 & 34.47932 & 0.02901 & 10.60 & 1.03 & -0.68 & -0.41 & 39.23 & inflow \\
8726 & 3703 & 116.92145 & 22.78161 & 0.02789 & 10.49 & 0.55 & -0.69 & -0.41 & 28.92 & inflow \\
8952 & 12705 & 206.03612 & 26.34891 & 0.06507 & 10.86 & 0.74 & -0.55 & -0.40 & 48.67 & inflow \\
9095 & 9102 & 243.08489 & 23.00202 & 0.03226 & 10.65 & 0.49 & -0.55 & -0.40 & 43.45 & inflow \\
9491 & 1902 & 120.04224 & 19.23433 & 0.07821 & 11.08 & 0.85 & -0.53 & -0.39 & 39.52 & inflow \\
9497 & 3703 & 117.99588 & 21.70878 & 0.04990 & 10.54 & 0.71 & -0.34 & -0.34 & 48.30 & inflow \\
9500 & 12702 & 132.59833 & 25.95408 & 0.02765 & 10.64 & 0.27 & -0.53 & -0.32 & 49.67 & inflow \\
9508 & 6103 & 127.34368 & 26.65689 & 0.05661 & 10.90 & 0.54 & -0.44 & -0.42 & 46.05 & inflow \\
9868 & 12705 & 218.87457 & 46.20239 & 0.07314 & 10.96 & 1.00 & -0.22 & -0.40 & 49.46 & inflow \\
\hline
\end{longtable}
\twocolumn


\bsp  
\label{lastpage}
\end{document}